# Clustering in the 1.2 Jy $IRAS$ Galaxy Redshift Survey II: Redshift Distortions and $\xi(r_p, \pi)$ ⋆

Karl B. Fisher[1], Marc Davis[2], Michael A. Strauss[3], Amos Yahil[4], & John P. Huchra[5]

[1] *Institute of Astronomy, Madingley Rd., Cambridge, CB3 0HA, England*
[2] *Astronomy and Physics Departments, University of California, Berkeley, CA 94720*
[3] *Institute for Advanced Study, School of Natural Sciences, Princeton, NJ 08540*
[4] *Astronomy Program, State University of New York, Stony Brook, NY 11794-2100*
[5] *Center for Astrophysics, 60 Garden St., Cambridge, MA 02138*

**ABSTRACT**

We examine the effect of redshift space distortions on the galaxy two-point correlation function $\xi(r_p, \pi)$ as a function of separations parallel ($r_p$) and perpendicular ($\pi$) to the line of sight. Modeling $\xi(r_p, \pi)$ measured from a full-sky redshift survey of $IRAS$ galaxies allows us to characterize the moments of the velocity distribution function of pairs of galaxies. We are guided in our parameterization of models by results from numerical simulations of Cold Dark Matter (CDM) models.

It is essential that the model for $\xi(r_p, \pi)$ contain the effects of both the first and second moments of the velocity distribution function, as they distort the redshift space correlations in opposing directions. We develop a method of fitting for the amplitudes of the velocity moments, and show that we can recover the correct values in Monte-Carlo realizations of the data drawn from $N$-body simulations. We find that the relative velocity dispersion of pairs of $IRAS$ galaxies is $\sigma(r) = 317^{+40}_{-49}$ km s$^{-1}$ at $r = 1\,h^{-1}$Mpc, consistent with previous estimates derived from optically selected galaxy catalogues. Unfortunately, the use of this result to estimate $\Omega$ via the Cosmic Virial Theorem is thwarted by large systematic uncertainties, making the application of this theorem to existing redshift surveys of little value.

We also fit for the mean relative streaming velocity of pairs, $v_{12}(r)$, which describes the growth of fluctuations on both linear and nonlinear scales. We find that $v_{12}(r) = 167^{+99}_{-67}$ km s$^{-1}$ at $r = 4\,h^{-1}$Mpc, so that on average, approximately half the Hubble expansion velocity of pairs at this separation is canceled by infall. At $r = 10\,h^{-1}$Mpc, the amplitude of the streaming is lower and $v_{12}(r) = 109^{+64}_{-47}$ km s$^{-1}$. Linear perturbation theory then implies that $\Omega^{0.6}/b = 0.45^{+0.27}_{-0.18}$ on scales $\sim 10 - 15\,h^{-1}$Mpc. The amplitude of $v_{12}(r)$ is sensitive to the assumed shape of $\sigma(r)$; if the latter deviates substantially from a virialized form on small scale, our best fit amplitude of $v_{12}(r)$ can deviate by a factor of two. Our derived result for $\beta$ is intermediate between that found on $\sim 1\,h^{-1}$Mpc scales and on $15 - 40\,h^{-1}$Mpc scales, arguing either for a strong dependence of $\beta$ on scale, or an error in the determination of $\beta$ on one or more scales.

# 1   INTRODUCTION

Redshift surveys of galaxies can provide statistical estimates of the departures from Hubble flow without the use of direct distance measurements of individual galaxies. The prominent elongations in the redshift distribution of cluster galaxies ("Fingers of God") are the most conspicuous examples of these "peculiar velocities", but the effect is detectable even in the absence of prominent clusters. In particular, the correlation function measured in redshift space $\xi(s)$ differs from that in real space $\xi(r)$ due to two effects: on small scales, $\xi(s)$ is *suppressed* due to the elongation of dense structures along the line of sight in redshift space, as mentioned above, while on large scales, coherent infall into overdense regions and outflow out of underdense regions *enhances* the correlations (Kaiser 1987; Fisher *et al.* 1993a). These effects can be seen in redshift survey data by plotting the correlations as a function of the components of the separation of a pair of galaxies in the plane of the sky ($r_p$) and along the line of sight ($\pi$). Statistically significant deviations of $\xi$ from concentric circles in the ($r_p, \pi$) plane are due to redshift space distortions: on small scales, we expect the contours to be elongated along the $\pi$ axis (a prolate distortion, considering $\xi(r_p, \pi)$ to have rotational symmetry about the $\pi$ axis), while on large scales, coherent infall into overdense regions elongates the contours along the $r_p$ axis, giving an oblate distortion.

Previous analyses of redshift space distortions from survey data (Rivolo & Yahil 1981; Davis & Peebles 1983; Bean *et al.* 1983; de Lapparent *et al.* 1988; Hale-Sutton *et al.* 1989; Mo *et al.* 1993) have been primarily concerned with the effect of virialized motions of close pairs. On larger scales, a number of authors have made linear theory predictions of the enhancement of the correlation function (Sargent & Turner 1977; Kaiser 1987; Lilje & Efstathiou 1989; McGill 1990; Hamilton 1992), which have been tested to some degree in $N$-body models (Lilje & Efstathiou 1989; Gramann, Cen, & Bahcall 1993). In particular, linear theory predicts that the angle-averaged correlation function $\xi(s)$ measured in redshift space is enhanced over the real space correlation function, $\xi(r)$, by a factor

$$K(\beta) = \left[1 + \frac{2}{3}\beta + \frac{1}{5}\beta^2\right] \quad , \tag{1}$$

where $\beta$ is related to the linear theory growth factor $D(t)$ by $\beta \equiv b^{-1} d\ln D/d\ln a \approx \Omega^{0.6}/b$, and $b$ is the linear bias parameter (assumed throughout this paper to be independent of scale) which relates the fluctuations of the galaxy counts to those in the underlying matter distribution, $(\Delta n/n) = b(\delta\rho/\rho)$. Hamilton (1993a) has detected the large scale infall using angular moments of $\xi(r_p, \pi)$ in a redshift survey of galaxies detected by the *Infrared Astronomical Satellite* (*IRAS*), flux-limited to 1.936 Jy at 60$\mu$m (Strauss *et al.* 1992b); he concludes that $\beta = 0.66^{+0.34}_{-0.22}$.

In a previous paper (Fisher *et al.* 1993b; hereafter Paper 1), we presented measurements of the amplitude of the correlation function $\xi(r_p, \pi)$ for a sample of 5313 *IRAS* galaxies flux-limited to 1.2 Jy at 60$\mu$m, selected from the *IRAS* database (Strauss *et al.* 1990; Fisher 1992). The sample covers 87.6% of the sky, and probes to an effective depth of $\approx$ 8000 km s$^{-1}$. In this paper, we focus on the redshift space distortions in $\xi(r_p, \pi)$, in order to characterize the amplitude of the velocity field of galaxies. The quantity $\xi(r_p, \pi)$ is related to $\xi(r)$ through the distribution function of the difference between the peculiar velocity of pairs of galaxies at a given separation, $F(\mathbf{w}|\mathbf{r})$. This quantity is of fundamental interest, because on large scales, linear theory relates the first moment of $F$ to the cosmological density parameter $\Omega$ and the bias parameter $b$. On small scales, the Cosmic Virial Theorem connects the second moment of $F$ to $\Omega$ and $b$. Thus in § 2, after presenting the *IRAS* $\xi(r_p, \pi)$ itself, we define two projections of $\xi(r_p, \pi)$ designed to be maximally sensitive to the two moments; we will fit models directly to these projections.

In § 3, we present the formalism needed for modeling $\xi(r_p, \pi)$, and show the predictions of linear theory and the Cosmic Virial Theorem. Both the shape of $F$ and the scaling of its moments with separation are unknown. Rather than trying to extract these unknown functions from the data, we turn to $N$-body simulations of several variants of Cold Dark Matter (CDM) models to characterize their forms (§ 4.1). In addition, we extract many Monte-Carlo realizations of the observational sample from one of the $N$-body simulations in order to characterize the errors in the observed quantities, thus allowing a $\chi^2$ fit of models for the redshift distortions to the data (§ 4.2). We minimize $\chi^2$ with respect to the unknown form of $F$ at a given separation, as well as the normalizations of the first and second moments of the velocity distribution function. We test our methods with the Monte-Carlo simulations, and show that we recover unbiased estimates of the normalizations of the velocity moments (§ 4.3). In § 5, we apply these techniques to the *IRAS* data themselves, and discuss their implications for $\Omega$ and $b$ in § 6. We conclude in 7.

# 2   THE OBSERVED QUANTITIES

## 2.1   The Observed $\xi(r_p, \pi)$

As described in Paper 1, we quantify the effects of the redshift space distortions by computing the correlation function as a function of separations parallel and perpendicular to the observer's line of sight. Explicitly, given a pair of galaxies with redshift positions, $\mathbf{v}_1$ and $\mathbf{v}_2$, we define a separation in redshift space, $\mathbf{s} = \mathbf{v}_1 - \mathbf{v}_2$, and an observer's line of sight, $\mathbf{l} = \frac{1}{2}(\mathbf{v}_1 + \mathbf{v}_2)$. We then define separations which are parallel ($\pi$) and perpendicular ($r_p$) to the line of sight,

$$\pi = \frac{\mathbf{s} \cdot \mathbf{l}}{|\mathbf{l}|} \quad , \quad (2)$$
$$r_p^2 = \mathbf{s} \cdot \mathbf{s} - \pi^2 \quad .$$

With these new variables, we can write down a generalized version of the estimator for the correlation function,

$$\xi(r_p, \pi) = \frac{N_{DD}(r_p, \pi) N_{RR}(r_p, \pi)}{N_{DR}^2(r_p, \pi)} - 1 \quad , \quad (3)$$

where $N_{DD}(r_p, \pi)$, $N_{DR}(r_p, \pi)$, and $N_{RR}(r_p, \pi)$ refer to data-data, data-random, and random-random pairs with separations $\pi$ and $r_p$ respectively. This estimator was suggested by Hamilton (1993b) as less sensitive to uncertainties in the mean density of the sample than is the traditional estimator used by Davis & Peebles (1983) and in Paper 1; in practice we obtained almost indistinguishable results using the two estimators. Minimum-variance weighting of the pairs is done, as described in Paper 1, with an effective depth between 50 and 110 $h^{-1}$ Mpc, depending on separation.

Figure 1a shows $\xi(r_p, \pi)$ for the full 1.2 Jy *IRAS* sample. The $\xi(r_p, \pi)$ in Figure 1a was computed using all the galaxies in the flux-limited sample in the redshift[†] range $500 < cz < 30,000$ km s$^{-1}$, with bin widths and separations of $1\, h^{-1}$Mpc. The contours are spaced at intervals of 0.1 dex above $\xi = 1$ (the heavy solid curve); for smaller values of $\xi$, the contours are spaced at arithmetic intervals of 0.1. The $\xi = 0$ contour is the heavy dashed curve, and negative contours are light dashes. The behavior of the redshift space correlation function, $\xi(s)$, is represented by dotted circular arcs with radii equal to the values of $s$ at which $\xi(s) = 1.0, 0.5$, and 0.25. For graphical clarity, the map has been twice smoothed by a 1-2-1 boxcar in each direction (although the model fitting below is done to the unsmoothed $\xi(r_p, \pi)$). The small scale distortion discussed above is evident as the stretching of the $\xi(r_p, \pi)$ contours along the $\pi$ direction for values of $r_p \lesssim 2\, h^{-1}$Mpc. The weak compression of the contours along the $\pi$ axis for $r_p \gtrsim 5\, h^{-1}$Mpc is the signature of the large scale redshift distortion.

In Paper 1, we projected $\xi(r_p, \pi)$ onto the $r_p$ axis to estimate the real space correlation function; modeling it with a power law yielded

$$\xi(r) = \left(\frac{r}{3.76^{+0.20}_{-0.23}\, h^{-1}\,\text{Mpc}}\right)^{1.66^{+0.12}_{-0.09}} \quad , \quad (4)$$

on scales $r \lesssim 20\, h^{-1}$ Mpc. We will use this result extensively below in the modeling of $\xi(r_p, \pi)$ itself.

Because the correlation function is a pair-weighted statistic, one expects it to be quite sensitive to the nature of the highest-density regions in the sample. Infrared-selected galaxies are known to give lower estimates of the galaxy density in cluster cores than do optically selected galaxies (Strauss *et al.* 1992a); as the cluster cores are regions with the highest pairwise velocity dispersion, this may cause us to underestimate the velocity dispersion. In order to address this, we calculated $\xi(r_p, \pi)$ for a "boosted" *IRAS* sample, in which galaxies in the cores of the seven clusters in Table 2 of Strauss *et al.* are given double weight. The results are shown in Figure 1b. Reassuringly, this figure differs little in Figure 1a; we will see that the two yield very similar estimates for the velocity dispersion on small scales (§5).

Sampling fluctuations are a major source of noise in the estimate of $\xi(r_p, \pi)$; there are a finite number of filaments and superclusters within the 1.2 Jy survey volume, with a finite number of orientations. Thus note the peaks in $\xi(r_p, \pi)$ along the $\pi$ axis at $\pi \approx 12, 17$, and $22\, h^{-1}$ Mpc, and the extension of the $\xi = 0.3, 0.4$, and 0.5 contours in the $\pi$ direction at $r_p \approx 8\, h^{-1}$Mpc, which presumably are due to particular structures in the sample which have not been averaged out. To gauge the reproducibility of the features of $\xi(r_p, \pi)$, we subdivide the *IRAS* sample into a North and South catalogue, split at Galactic latitude $b = 0$. Figures 1c and 1d show $\xi(r_p, \pi)$ for these two subcatalogues. Roughly speaking, the two subsamples are very similar for $\xi(r_p, \pi) > 0.4$; the discrepancies at smaller amplitudes are indicative of the noise of the derived $\xi(r_p, \pi)$ on large scales. The $\xi(r_p, \pi) = 1$ contour is considerably more prolate in the Northern than Southern catalogue, presumably due to the closer proximity of the clusters within the Northern hemisphere. Indeed, our model fits below indicate that the velocity dispersion generating this "Finger of God" is larger in the North than the South.

## 2.2 Projections of the $\xi(r_p, \pi)$ Plot

We will find it useful in fitting models to the data to work with two projections of the $\xi(r_p, \pi)$ contours, which are less affected by the finite volume of the survey than $\xi(r_p, \pi)$ itself. Our primary aim will be to constrain the first and second moments of the velocity distribution function, and thus we choose projections of $\xi(r_p, \pi)$ which have maximum sensitivity to each of these.

---

[†] All redshifts are corrected for the motion of the sun with respect to the barycenter of the Local Group, following Yahil, Tammann, & Sandage (1977); no further corrections are applied.

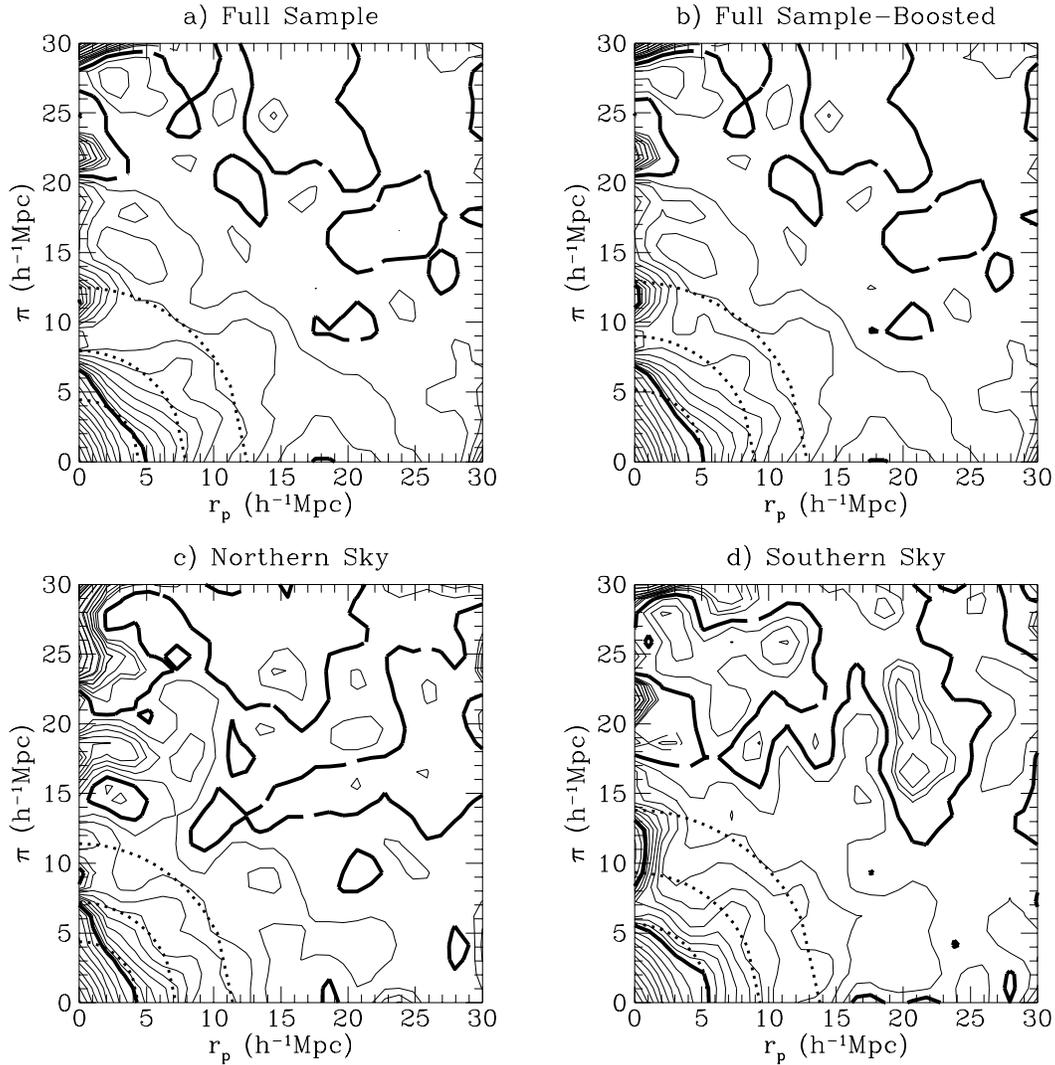

**Figure 1.** The *IRAS* 1.2 Jy $\xi(r_p, \pi)$ for the *a)* full sample, *b)* the full sample with clusters boosted, *c)* the Northern subsample ($b > 5°$), and *d)* the Southern subsample ($b < 5°$). All galaxies in the sample with redshifts between 500 and 30,000 km s$^{-1}$ are used. The contours are spaced at intervals of 0.1 dex above $\xi = 1$ (the heavy solid curve); for smaller values of $\xi$, the contours are spaced at arithmetic intervals of 0.1. The $\xi = 0$ contour is the heavy dashed curve, and negative contours are light dashes. The behavior of the redshift space correlation function, $\xi(s)$, is represented by dotted circular arcs with radii equal to the values of $s$ at which $\xi(s) = 1.0, 0.5$, and 0.25. For graphical clarity, the map has been twice smoothed by a 1-2-1 boxcar in each direction. Note that on small scales, the contours are distorted in the $\pi$ direction, while on larger scales, a distortion in the $r_p$ direction is evident.

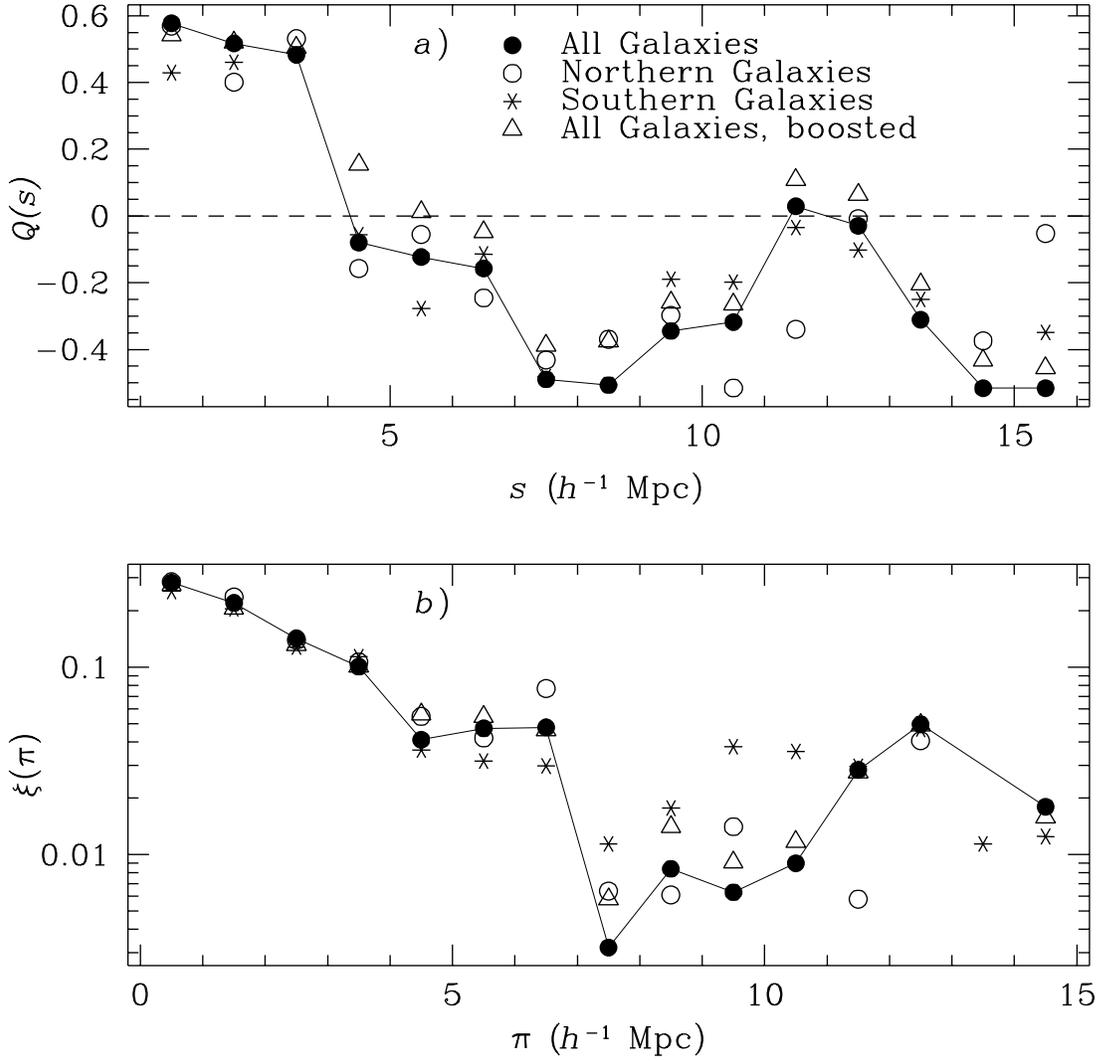

**Figure 2.** *a).* The normalized quadrupole moment $Q(s)$ for the full *IRAS* sample (filled circles), the Northern (open circles) and Southern (stars) subsamples, and the full sky with clusters boosted (triangles). Note that $Q(s)$ becomes negative for $s > 4\,h^{-1}$ Mpc. *b).* Same as in *a)* for $\xi(\pi)$.

Hamilton (1992) has emphasized the use of multipole moments of the $\xi(r_p, \pi)$ maps to measure the streaming on large scales. We shall here limit ourselves to consideration of the quadrupole, which is the major distortion expected as we will see below (Equation 23). Consider the statistic $Q(s)$ defined by

$$Q(s) \equiv 5\,\frac{\int\limits_0^1 d\mu\,\mathcal{P}_2(\mu)\xi(r_p,\pi)}{\int\limits_0^1 d\mu\,\mathcal{P}_0(\mu)\xi(r_p,\pi)}\,, \qquad (5)$$

where $\mu = \pi/s$, and $\mathcal{P}_n(\mu)$ is the Legendre polynomial of order $n$. The $Q(s)$ defined here is the quantity $\xi_2(s)$ defined by Hamilton (1992), normalized by the zeroth moment. We apply this normalization in order to minimize sensitivity to the amplitude of $\xi(r_p,\pi)$, and thus maximize the sensitivity to the redshift distortions themselves.

Figure 2a shows the $Q(s)$ statistic, for the full sky analysis (solid connected points), the North and South subsamples (open circles and stars, respectively), and the boosted sample (triangles). Note that $Q(s)$ is positive for small $s$, but becomes negative, corresponding to an oblate distortion of $\xi(r_p,\pi)$ along the $r_p$ axis, for $s > 4\,h^{-1}$ Mpc. Moreover, the $Q(s)$ for the full sky, the two subsamples, and the boosted sky agree quite well. $Q(s)$ does not asymptote to zero at large scales, because of the normalizing factor, although we will see below that the signal to noise ratio in this statistic drops below unity on scales above $\sim 10\,h^{-1}$ Mpc. Unfortunately, even on scales as large as 15 $h^{-1}$ Mpc, $Q(s)$ is quite sensitive to the value of the dispersion on small scales (cf., Figure 7) and therefore we are forced into a simultaneous fit for both the dispersion and the streaming.

On small scales, the clustering is presumed to be stable in the sense that the average separation of galaxy pairs is constant in physical coordinates; consequently the first moment of the velocity distribution function (streaming motion) approaches zero for small pair separations. Thus in order to isolate the effect of small-scale velocity dispersion (the second moment), we look at a statistic with most of its weight on small scales. With perfect data, it would be ideal to take a cut through the contour plot at $r_p = 0$, but in practice, this is very noisy. Thus we sum over the first two bins in $r_p$ to define:

$$\langle \xi(r_p = 1, \pi) \rangle \equiv 0.5\,[\xi(r_p = 0.5\,h^{-1}\,\mathrm{Mpc}, \pi) + \xi(r_p = 1.5\,h^{-1}\,\mathrm{Mpc}, \pi)] \quad . \tag{6}$$

As with $Q(s)$, we are mainly interested in fitting the *shape* of this function in order to constrain the velocity dispersion; we wish our fitting to be insensitive to the amplitude of $\xi(r_p, \pi)$. Therefore, in analogy with the normalization of $Q(s)$, we normalize the statistic in Equation 6,

$$\xi(\pi) \equiv \frac{\langle \xi(r_p = 1, \pi) \rangle}{\int_0^{\pi_{max}} \langle \xi(r_p = 1, \pi) \rangle\,d\pi} \quad , \tag{7}$$

where $\pi_{max}$ is the maximum value of $\pi$ used in the analysis; in the results presented below $\pi_{max} = 15\,h^{-1}$ Mpc.

The solid connected points in Figure 2b plot $\xi(\pi)$ for the full sample, while the open circles and stars are for the Northern and Southern portions, respectively. The three sets of points agree with one another very well, implying that this statistic is more robust than $\xi(r_p, \pi)$ as a whole. The quantity $\xi(\pi)$ is almost a straight line in this linear-log plot, and is positive to $\sim 10\,h^{-1}$ Mpc, beyond which the errors (which we will quantify below) dominate.

## 3   MODELING $\xi(r_p, \pi)$

In Paper 1, we projected the contour maps of $\xi(r_p, \pi)$ onto the $r_p$ axis to extract the underlying real space correlation function $\xi(r)$ (Equation 4). However, we can extract more information from our plot of $\xi(r_p, \pi)$ than $\xi(r)$ alone, as the distortions in redshift space contain information on the velocity distribution function of pairs of galaxies. Let $F(\mathbf{w}|\mathbf{r})$ be the distribution function of velocity differences $\mathbf{w}$ for pairs of galaxies separated by vector distance $\mathbf{r}$. A convenient model for $\xi(r_p, \pi)$ is (Peebles 1980, Equation 76.6),

$$1 + \xi(r_p, \pi) = \int d^3\mathbf{w}\,F(\mathbf{w}|\mathbf{r})\,[1 + \xi(r)] \quad , \tag{8}$$

where $r^2 = r_p^2 + (\pi - w_3/H_o)^2$, $H_o$ is the Hubble constant, and $w_3$ is the component of $\mathbf{w}$ along the line of sight. Thus $\xi(r_p, \pi)$ may be thought of as a convolution of the real space correlation function with the velocity distribution function. This expression does not take into account the correlations one expects between particular features in the density and velocity fields (e.g., coherent infall into overdense regions). However, because $\xi(r_p, \pi)$ is computed by averaging over all environments, such effects should average out to the extent to which our survey volume represents a "fair" sample of the universe.

A good approximation to Equation 8 in the regime of slowly varying velocity dispersion is (Peebles 1980, Equation 76.12)

$$1 + \xi(r_p, \pi) = \int dw_3\,f(w_3|r)\left\{1 + \xi\left[\left(r_p^2 + (\pi - w_3/H_o)^2\right)^{1/2}\right]\right\} \quad , \tag{9}$$

where $f(w_3|r)$ is the velocity distribution function averaged over directions perpendicular to the line of sight,

$$f(w_3|r) \equiv \int dw_1\,dw_2\,F(\mathbf{w}|\mathbf{r}) \quad . \tag{10}$$

If we wish to fit Equation 9 to the data, we need a model for the distribution function $f(w_3|r)$. Let us start by parameterizing $F(\mathbf{w}|\mathbf{r})$ in terms of its first and second moments; as we will see below, it is these moments which will allow us to draw conclusions about $\Omega$. The first moment, the mean relative velocity of galaxy pairs of separation $\mathbf{r}$, we label $v_{12}(\mathbf{r}) \equiv -\langle[\mathbf{v}(\mathbf{x}) - \mathbf{v}(\mathbf{x} + \mathbf{r})] \cdot \hat{\mathbf{r}}\rangle$, where the average is number weighted, not volume-weighted. By symmetry this mean streaming must be directed along the separation vector of the pair, and is thus a function only of the magnitude of the vector $\mathbf{r}$; we shall adopt a convention that assigns positive streaming for infall, in opposition to Hubble expansion. The corresponding first moment of $f$ is $\langle w_3 \rangle = y\,v_{12}(r)/r$, where $y$ is the component of $\mathbf{r}$ in the line-of-sight direction. The second moment of $F$ is the 3-d relative velocity dispersion of pairs, $\Sigma(r) \equiv \langle(\mathbf{v}(\mathbf{x}) - \mathbf{v}(\mathbf{x} + \mathbf{r}) + v_{12}(r)\hat{\mathbf{r}})^2\rangle^{1/2}$. In general, we should distinguish between the longitudinal and transverse velocity dispersion (cf., § 4.2); for now, however, we assume an isotropic velocity dispersion tensor. Under this approximation of isotropy, the second moment of $f$ is just the 1-d dispersion, $\sigma^2(r) = \Sigma^2(r)/3$. These moments figure prominently in the kinetic theory formulation of clustering dynamics (Davis & Peebles 1977; Peebles 1980) and can be readily measured in $N$-body simulations. Their qualitative behavior is insensitive to the power spectrum, as we will see in § 4.1 below.

We parameterize the functional dependence of $f(w_3|r)$ on its first and second moments by writing

$$f(w_3|r) = \frac{1}{\sigma(r)} g\left(\frac{w_3 - \langle w_3(r) \rangle}{\sigma(r)}\right) \quad . \tag{11}$$

In particular, consider the family of models in which $g$ is an exponential of a power of its argument. Let the vector $r$ separating two galaxies be decomposed into a transverse component $r_p$ and a line-of-sight component $y$. Then $w_3 = \pi - y$ and Equation 9 gives

$$1 + \xi(r_p, \pi) = C \int \frac{dy}{\sigma(r)} (1 + \xi(r)) \exp\left[-\eta \left|\frac{\pi - y - y\frac{v_{12}(r)}{r}}{\sigma(r)}\right|^\nu\right] \quad , \tag{12}$$

where $r^2 = r_p^2 + y^2$. The normalization constants $C$ and $\eta$ are given by

$$\eta = \left[\frac{\Gamma\left(\frac{3}{\nu}\right)}{\Gamma\left(\frac{1}{\nu}\right)}\right]^{\nu/2} \quad \text{and} \quad C = \frac{\nu \eta^{1/\nu}}{2 \Gamma(\frac{1}{\nu})} \quad . \tag{13}$$

We will see below that the exponential model $\nu = 1$ provides an adequate fit for both the data and for $N$-body models; in this case, $\eta = \sqrt{2}$ and $C = 1/\sqrt{2}$.

The role of $v_{12}(r)$ can be best understood by examining the first momentum moment of the second BBGKY equation, which expresses the conservation of particle pairs (Davis & Peebles 1977; Peebles 1980, Equation 71.6),

$$\frac{\partial \xi}{\partial t} - \frac{1}{a^2 \, x^2} \frac{\partial}{\partial x}\left[x^2 (1 + \xi) v_{12}\right] = 0 \quad , \tag{14}$$

where $x$ is a comoving length, $a$ is the scale factor, and $v_{12}(x,t)$ is the *proper*, not comoving, peculiar velocity.

Let us solve for $v_{12}$ of the *dark matter* on large scales; the biasing of galaxies enters in a subtle way, as we shall see. In the limit $\xi \ll 1$,

$$\int_0^x dx' x'^2 \xi(x',t) \propto D^2(t) \quad , \tag{15}$$

where $D(t)$ is the linear growth rate of dark matter fluctuations appropriate for a given cosmology. Equations 14 and 15 yield

$$v_{12}(x,t) = \frac{2}{3} f(\Omega) a^2 H_0 \, x \, \tilde{\xi}(x) \quad , \tag{16}$$

where $f(\Omega) = d\ln D/d\ln a$ and

$$\tilde{\xi}(x) \equiv \frac{3}{x^3} \int_0^x dx' \, x'^2 \xi(x') \quad . \tag{17}$$

This equation relates the streaming and correlation function of dark matter, while what is observed are galaxies. The quantity $v_{12}$ is a density weighted mean of the pairwise velocities (cf., the discussion in Bertschinger 1992). That is, if expressed in terms of a volume-weighted mean:

$$v_{12}(r) = \langle -\left[(\mathbf{v}(\mathbf{x}) - \mathbf{v}(\mathbf{x} + \mathbf{r})) \cdot \hat{\mathbf{r}}\right] [1 + \delta(\mathbf{x})][1 + \delta(\mathbf{x} + \mathbf{r})]\rangle_{volume} \quad . \tag{18}$$

We can expand this into a term independent of $\delta$, a term proportional to $\delta$, and a term proportional to $\delta^2$. The latter term can be dropped in the limit $\delta \ll 1$. In linear theory, the volume-weighted pairwise velocity field is symmetric, and thus the zeroth order term drops out. Thus

$$v_{12}(r) = \langle [-(\mathbf{v}(\mathbf{x}) - \mathbf{v}(\mathbf{x} + \mathbf{r}) \cdot \hat{\mathbf{r}})] [\delta(\mathbf{x}) + \delta(\mathbf{x} + \mathbf{r})]\rangle_{volume} \quad . \tag{19}$$

Because the velocity field $\mathbf{v}$ is the same for dark matter and galaxies, in a biasing model in which $\delta_{galaxies} = b\delta_{dark\ matter}$, we have $v_{12\ galaxies} = b v_{12\ dark\ matter}$. Similarly, $\xi_{galaxies} = b^2 \xi_{dark\ matter}$, and thus Equation 16 yields

$$v_{12\ galaxies}(x) = \frac{2}{3} \beta a^2 H_0 \, x \, \tilde{\xi}_{galaxies}(x) \quad , \tag{20}$$

where $\beta = f(\Omega)/b$. A slightly more accurate approximation would be to write:

$$v_{12\ galaxies}(x) = \frac{2}{3} \beta a^2 H_0 \, x \, \tilde{\xi}_{galaxies}(x) / (1 + \xi(x)) \quad , \tag{21}$$

and indeed, this is what we shall use in this paper. For convenience, in the remainder of this paper we will drop the subscript *galaxies* on $v_{12}(r)$ and $\xi(r)$, bearing in mind that these quantities refer to the galaxy distribution and not the mass.

What is the signature of large-scale streaming in $\xi(r_p, \pi)$? If the dispersion is a slowly varying function of scale (a good approximation on large scales, as we shall see in § 4.1) and $\xi \ll 1$, then Equation 12 can be expanded in $v_{12}(r)/H_0 r$ to give

$$\xi(r_p, \pi) = \xi(s) - \frac{v_{12}(s)}{H_0 s}\left[1 + s\mu^2 \frac{d}{ds}\ln\frac{|v_{12}(s)|}{s}\right] + \mathcal{O}\left(\frac{v_{12}}{H_0 s}\right)^2 \quad , \tag{22}$$

where $s^2 = r_p^2 + \pi^2$, $\mu = \pi/s$, and $\xi(s)$ is the direction-averaged redshift space correlation function (cf., Peebles 1980, equation 76.9). In the linear regime where $v_{12}(r)$ is given by Equation 20, Equation 22 can be written as

$$\xi(r_p, \pi) = \left(1 + \frac{2}{3}\beta\right)\xi(s) + \frac{4}{3}\beta\mathcal{P}_2(\mu)\left[\xi(s) - \tilde{\xi}(s)\right] \quad . \tag{23}$$

Thus the dominant effect of the streaming is to introduce a quadrupole moment to $\xi(r_p, \pi)$ (which is the motivation for defining $Q(s)$ as we did in the previous section). Although our expansion of $\xi(r_p, \pi)$ to first order in $v_{12}(r)/H_0 r$ is *not* equivalent to a first order expansion in the density field $\delta$, Equation 23 agrees to first order in $\beta$ with the result of Hamilton (1992) (his linear theory expansion also includes a hexadecapole term). In particular, the angle average of Equation 23 gives $\langle\xi(r_p, \pi)\rangle_\mu = (1 + (2/3)\beta)\xi(r)$, which is Kaiser's (1987) result to first order in $\beta$ (Equation 1). The actual linear theory limit of Equation 12 differs from Equation 22 because we have ignored the scale dependence of the linear theory dispersion here. However, our modeling of $\xi(r_p, \pi)$ below takes this scale dependence into account explicitly, meaning that our results are consistent with linear theory on large scales (cf., Figure 4).

Equation 21 points us towards our first goal; if our model fits to $\xi(r_p, \pi)$ can give us $v_{12}(r)$, we can put constraints on $\beta$. Another constraint on $\Omega$ and $b$ comes from the second moment, $\sigma(r)$. In the continuum limit, the statistical equilibrium of clustering on small scales implies that the tidal gravitational acceleration acting on pairs of galaxies is balanced by the kinetic pressure of their relative dispersion (Peebles 1980). Modeling the two-point correlation function as a power law of index $\gamma$ and the three-point correlation function as a power law of index $2\gamma$ yields the Cosmic Virial Theorem:

$$\Sigma^2(r) - v_{12}^2(r) = \frac{9\,\Omega\,Q(H_0 r)^2 \xi(r) J(\gamma)}{4\,b\,(\gamma-1)(2-\gamma)(4-\gamma)} \quad , \tag{24}$$

where $Q$ (not to be confused with $Q(s)$ defined in the previous section!) expresses the proportionality between the two- and three-point correlation functions and $J(\gamma)$ is the dimensionless integral of Peebles (1980, Equation 75.13). The poles in Equation 24 for $\gamma = 1$ and $\gamma = 2$ attest to the sensitivity of the result to the power law form of the three point correlations. This expression is derived assuming that the velocity dispersion tensor is isotropic, and thus the three-dimensional velocity dispersion $\Sigma$ is $3^{1/2}\sigma$. The dependence of the Cosmic Virial Theorem on biasing is ambiguous. As discussed by Bartlett & Blanchard (1993), the velocity dispersion of *galaxies* depends on the ratio of the galaxy-galaxy-mass three point correlation function to the galaxy-galaxy two-point correlation function. If we assume linear bias, Equation 24 follows. However, the Cosmic Virial Theorem is valid on small scales where the density contrasts are much larger than unity and linear bias is unlikely to be applicable. Furthermore, the bias appearing in Equation 21 is on large scales where $\xi \ll 1$, and there is no guarantee that it has any direct relation to the small-scale bias. Indeed, Bartlett & Blanchard (1993) show that the interpretation of the quantity $\Omega/b$ which we solve for using Equation 24 in fact depends sensitively on the biasing model assumed. However, we show in § 6 that the combination of statistical and systematic uncertainties entering into the Cosmic Virial Theorem make it almost useless as an estimator of cosmological parameters, and thus the ambiguity of the role of biasing becomes a moot issue.

## 4 CALIBRATION USING $N$-BODY SIMULATIONS

In order to model the observed $\xi(r_p, \pi)$ (or equivalently, its projections defined in § 2.2) from Equation 12, several ingredients are needed. The real space correlation function of *IRAS* galaxies is known (Equation 4), but the velocity distribution function as a function of separation is not. Our approach is to measure the distribution function and its moments from a series of $N$-body models, and use this directly in our modeling, varying only the form of the distribution function and the amplitude of its moments. This approach is valid to the extent to which the derived best amplitudes are insensitive to the details of the velocity moments adopted. The $N$-body simulations themselves, and their velocity moments, are presented in § 4.1. We also extract mock *IRAS* redshift surveys from one of the $N$-body simulations, from which we measure $Q(s)$ and $\xi(\pi)$, and thus quantify the covariance matrix between the observational quantities (§ 4.2). This allows us to fit models to the data by minimizing $\chi^2$; in § 4.3, we test this by applying it to the mock redshift surveys themselves, and demonstrate that we recover unbiased estimates of the amplitudes of the velocity moments. We emphasize that we are *not* using these $N$-body simulations to test the models that underlie the simulations; rather, we use them as a bridge over our ignorance of the true form of the velocity distribution function, for our modeling of $\xi(r_p, \pi)$ and its errors, and in order to test our fitting methods.

### 4.1 Velocity Moments

We have derived velocity moments from three variants on the Cold Dark Matter (CDM) model of structure formation. We will see that the *shape* (but not the normalization) of the moments as a function of $r$ varies little from model to model, which suggests that they are generally applicable (Efstathiou *et al.* 1988) and can serve as templates for our modeling.

The first simulation employed is that of a standard CDM model ($\Omega = 1$, $\Omega h = 0.5$ and $\Omega_\Lambda = 0$; hereafter SCDM); the details of this simulation are given in Frenk *et al.* (1990). We compute the correlation function and velocity moments of both unbiased and biased subsets of the particles. The unbiased statistics are derived from a random fraction of the 262,144 simulation particles at the time when the correlation amplitude of the points is close to that of the *IRAS* galaxies; we designate this epoch as redshift $z = 0$. The correlation statistics in the biased model (hereafter BCDM) are computed by weighting each simulation particle by the expected number of peaks per particle, using the "peak-background split" procedure of Bardeen *et al.* (1986). Galaxies are associated with $1.5\,\sigma$ peaks at an output time corresponding to $z = 0.6$, using the methods described in White *et al.* (1987). The correlation functions $\xi(r)$ of the BCDM and SCDM models are very similar, so the effective bias of the BCDM model is $b = (1 + z) = 1.6$. We also examined the velocity moments of a simulation of an unbiased open CDM model ($\Omega h = 0.14$, $\Omega_\Lambda = 0$; hereafter OCDM), kindly provided to us by G. Kauffmann from the model "M" simulation of Kauffmann & White (1992). We will measure the moments of the velocity distribution function from all three $N$-body models, but we measure the shape of $F$ and the covariance matrix of $Q(s)$ and $\xi(\pi)$, and test our fitting method, using the SCDM model alone.

A common problem with collisionless $N$-body simulations is that their small-scale velocity dispersions are much higher than indicated by observations, including those presented in this paper (e.g., Davis *et al.* 1985; Gelb & Bertschinger 1993). This is often taken as an argument against unbiased $\Omega = 1$ models, although including the non-gravitational physics of baryonic material might substantially decrease the velocity dispersions of pairs (however, see Cen & Ostriker 1992). In any case, we wish to produce models that closely mimic the data, and thus we convolve the velocity field of the points in the simulations with a Gaussian window with a width of $0.25\,h^{-1}$ Mpc. That is, our aim is not to test the underlying assumptions that go into generating the $N$-body models, but simply to obtain generic fitting functions for the velocity moments.

In applying these moments to the model fitting of *IRAS* data, there are three modifications we would like to make, in order that they more closely match what we expect given the clustering properties of *IRAS* galaxies, and to bring the curves into closer agreement with one another. First, the correlation functions of the *IRAS* galaxies and the $N$-body particles are not the same; in particular, they have slightly different correlation lengths ($r_0$=3.35, 4.3, and 5.0 $h^{-1}$ Mpc for the SCDM, BCDM, and OCDM models respectively, while the *IRAS* correlation length is $r_\circ =3.76\,h^{-1}$ Mpc). Thus we scale both the $r$ coordinate and amplitude of the velocity moments by the ratio of the correlation lengths. In addition, the correlation functions of the $N$-body models are somewhat steeper than that of *IRAS* galaxies, which will influence the small scale behavior of $\sigma(r)$. The Cosmic Virial Theorem (Equation 24) tells us that the dispersion of bound objects scales like

$$\sigma \propto (\delta M/r)^{1/2} \propto \left(r^2 \xi(r)\right)^{1/2} \propto r^{1-\gamma/2} \propto r^{0.17} \quad , \tag{25}$$

where the last proportionality holds for the *IRAS* value of $\gamma = 1.66$ (Paper I). We therefore replace the small scale velocity dispersion calculated from the $N$-body models on scales $r < r_\circ = 3.76\,h^{-1}$ Mpc with the scaling given in Equation 25; we furthermore hold $\sigma$ constant for $r < 1\,h^{-1}$ Mpc, as we find we get more stable estimates of the normalization of $v_{12}(r)$ when we do this.

Finally, on large scales, Equation 21 shows us the scaling of $v_{12}$ expected in linear theory. Because the $N$-body models and the *IRAS* galaxies have different correlation functions, the slope of $v_{12}(r)$ will differ. We thus modify the $N$-body $v_{12}(r)$ as follows:

$$v_{12}(r) \rightarrow v_{12}(r) \left[ \frac{1 + \left(\frac{r}{r_c}\right)^4 \left[2\beta r \tilde{\xi}(r)\right] / \left[v_{12}(r)(3 - \gamma)(1 + \xi(r))\right]}{1 + \left(\frac{r}{r_c}\right)^4} \right] \quad , \tag{26}$$

where $\xi$ and $\gamma$ are appropriate for the *IRAS* galaxies (Equation 4), $\beta$ is the value appropriate for the $N$-body model in question, $r_c = 5\,h^{-1}$ Mpc, and the form $\left(\frac{r}{r_c}\right)^4$ is an arbitrary but convenient form to force a sharp transition to the correct linear theory slope on scales above $r_c$.

Figure 3 plots $v_{12}(r)$ and $\sigma(r)$ as measured directly from the three simulations. The Gaussian smoothing of the velocity field is evident in the suppression of $\sigma$ on small scales $\lesssim 1\,h^{-1}$ Mpc. The left two panels show the velocity moments as measured from the simulations directly, while the right panels have had the *IRAS* rescaling of lengths and the further modifications of Equations 25 and 26 applied. With these modifications, although the three curves in each case have very different *amplitudes*, the shapes are similar.

The unbiased SCDM model has the largest amplitude velocity field, as expected. Note that $\sigma(r)$ is essentially flat on large scales for all models, and indeed, when the *IRAS* virialization condition is applied on small scales, $\sigma(r)$ remains constant to first order on all scales. Note that $\sigma(r)$ is the *rms* velocity dispersion about the mean for all pairs of a given separation. Most galaxies forming pairs of large separation are in low and medium density environments, while a higher proportion of galaxies forming pairs on smaller scales are taken from high density, high dispersion, environments. This is why $\sigma$ peaks at a separation characteristic of clusters, $\sim 2 - 3\,h^{-1}\,{\rm Mpc}$. On larger scales the dispersion declines to its asymptotic value of $(2/3)^{1/2}$ of the *rms* value of the 3-d velocity dispersion.

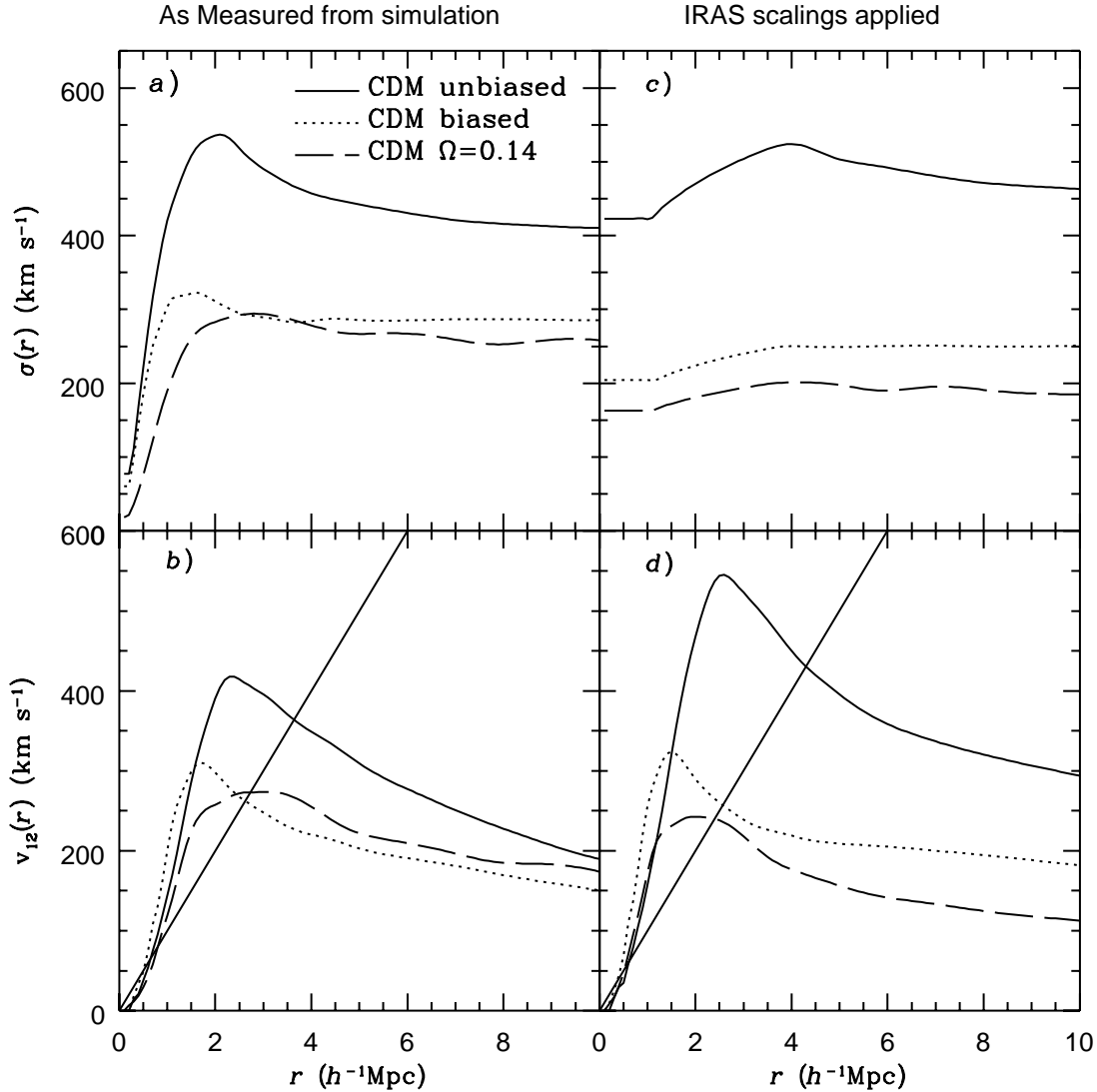

**Figure 3.** Velocity moments for the $N$-body simulations. The different line types give results for the different simulations used, as is indicated in the figure itself. *a)*. The one dimensional relative velocity dispersion, $\sigma(r)$. *b)* The mean pair-wise velocity, $v_{12}(r)$. The diagonal straight line in the Hubble line. *c)*. The quantity $\sigma(r)$, after the modifications to match the *IRAS* data on small and large scales, as described in the text. *d)*. The quantity $v_{12}(r)$, after the modifications to match the *IRAS* data.

The behavior of $v_{12}(r)$ is characteristic of that seen in all $N$-body experiments (e.g. Efstathiou *et al.* 1988): on large scales the mean infall streaming does not fully compensate for the Hubble expansion (the straight line of Figure 3*b* and *d*) but does retard the physical expansion of the pairs, while on smaller scales, $r \lesssim r_0$, the mean infall streaming overcompensates the Hubble flow, indicating that, on average, pairs on this scale are physically moving together.

The similarity of the shapes of these curves to one another means that we will not be able to distinguish observationally between them; we will only be sensitive to the amplitude. In particular, we will not readily be able to distinguish biased and unbiased models with the same values of $\beta \equiv \Omega^{0.6}/b$ from the velocity moments themselves. This is illustrated directly in Figure 4, which shows the quantity $3/2\,v_{12}(r)(1 + \xi(r))/(H_\circ\,r\,\tilde{\xi}(r))$ versus $r$, which, according to Equation 21, should approach the constant value of $\beta$ on large scales. Indeed, in each case, the curves approach the correct value of $\beta$ on scales above $7\,h^{-1}\,{\rm Mpc}$ or so. This is small enough that we have a reasonable chance of detecting the linear theory limit in $v_{12}$ in the *IRAS* data itself.

Thus far we have considered the *moments* of the velocity distribution function. What is the best model for its *shape*?

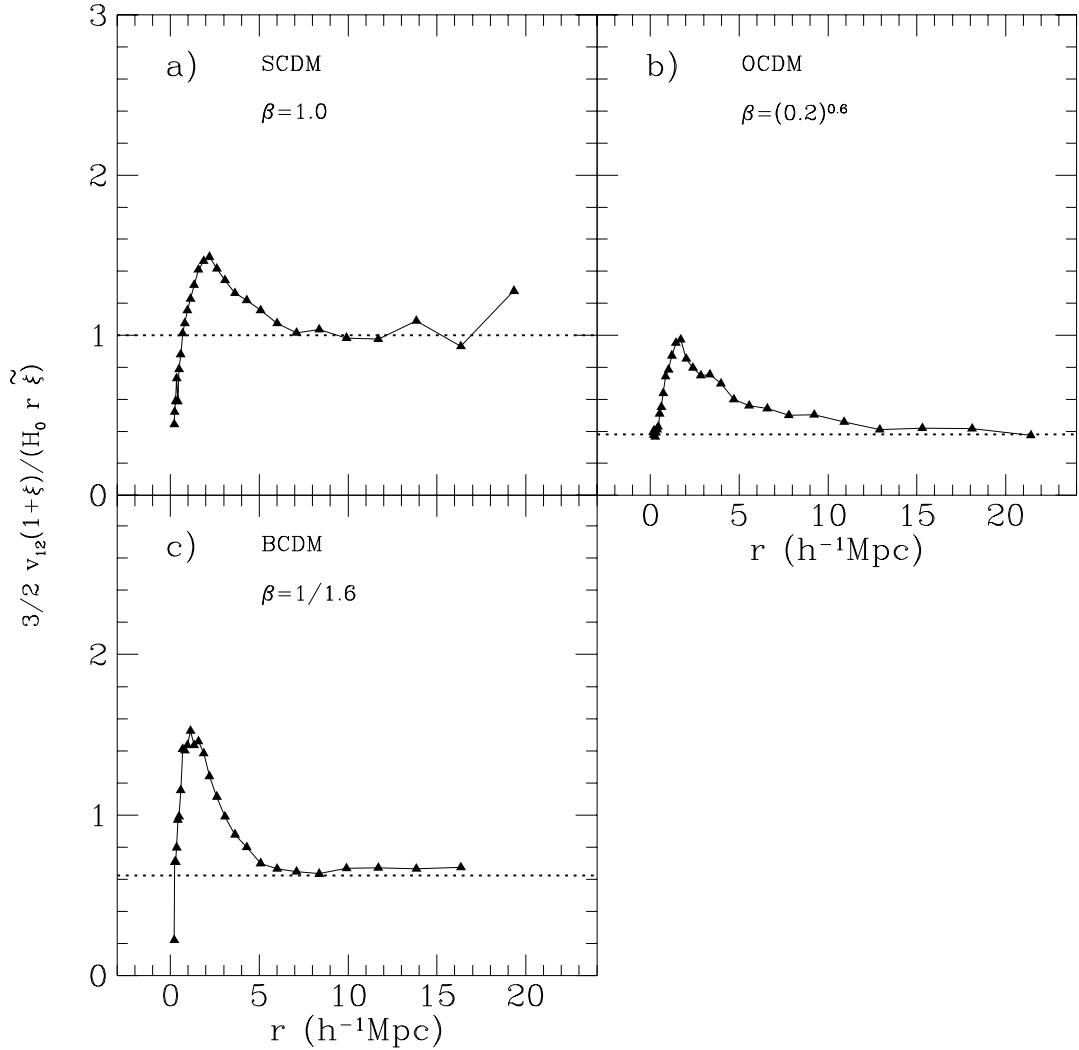

**Figure 4.** Plots of $v_{12}(r)(1 + \xi(r))/((2/3)H_0 r \tilde{\xi}(r))$ versus $r$ for the SCDM, BCDM, and OCDM models. In linear theory (large scales) this curve should take on the constant value $\beta = f(\Omega)/b$.

We consider the SCDM model in further detail to address this question. Figure 5a shows the longitudinal component of the relative velocity distribution function of pairs $f(w_\parallel|r)$ extracted from this $N$-body model, for various separations $r$, labeled by the amplitude of the clustering strength, $\xi(r)$. This is part of the underlying distribution from which $v_{12}(r)$ and $\sigma(r)$ of Figure 3 were computed. The distribution is quite skew-negative on all scales. The distribution function appears to have exponential tails, but has a softer than exponential core. If these curves are functions only of their first and second moments, then they will all coincide when scaled to a common mean and variance (Equation 11). Figure 5b applies this scaling, and indeed shows that they are close to being self-similar. The heavy solid line is a spline fit to the mean of these curves; this mean, together with the measured $v_{12}(r)$ and $\sigma(r)$, gives a reasonable approximation to the true $f(w_\parallel|r)$ for the $N$-body models. The transverse component of this distribution, $f(w_\perp|r)$, cannot, by symmetry, be skew, and the projected distribution, $f(w_3|r)$, is a combination of the two. Thus models using this spline form for $f(w_3)$ overestimate the skewness; our modeling below ignores this subtlety.

Our aim is to measure $\beta$ from the measured amplitude of $v_{12}(r)$ using the linear theory prediction of Equation 21 above. Unfortunately, on all scales probed by the present data set, the effect of $v_{12}(r)$ on $\xi(r_p, \pi)$ is strongly counteracted by that of $\sigma(r)$. That is, we cannot determine $v_{12}(r)$ without also determining $\sigma(r)$, so we must simultaneously model both the linear and non-linear aspects of the problem. This is illustrated in Figures 6a, b, and c, which show models of $\xi(r_p, \pi)$ from Equation 12 using an exponential velocity distribution and the SCDM model for $\xi(r)$ and the velocity moments. Figure 6a uses the SCDM velocity moments as plotted on the left side of Figure 3. In Figure 6*b*, we have doubled $\sigma(r)$ from that in Figure 3*a*, and in Figure 6*c* we have halved $\sigma(r)$ from that in Figure 3*a*. Figure 6*a* shows considerable oblateness on large scale and moderate prolateness on small scale, Figure 6*b* is quite prolate on all scales, while Figure 6*c* shows extreme oblateness, especially on large scale. Figure 7*a* shows the derived $Q(s)$ for these three maps. The $Q(s)$ for the low dispersion model goes very negative (oblate) on large scales, while the high dispersion model has positive $Q$ (prolate) on nearly all scales. Thus although in linear theory $Q(s)$ is primarily a measure of the first moment $v_{12}$ of the velocity distribution function, it is in fact strongly dependent on the amplitude of the velocity dispersion, which on small scales, $(r \lesssim 10 - 15\ h^{-1}\mathrm{Mpc})$, is dominated by nonlinear effects. Therefore estimates of $\beta$ from redshift distortions in $\xi(r_p, \pi)$, which have a measurable signal only for modest scale separations, must necessarily model the dispersion accurately on scales where linear theory is not applicable.

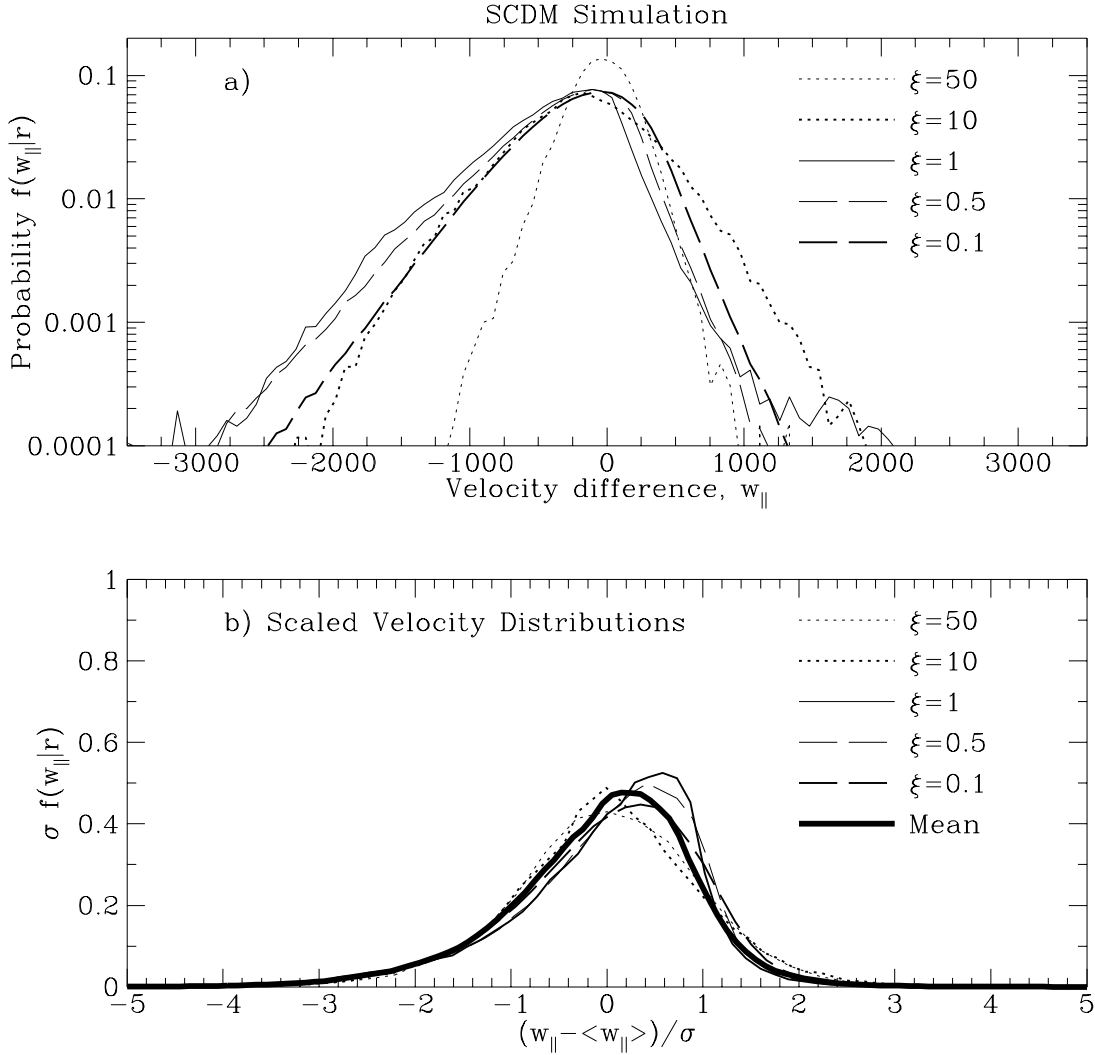

**Figure 5.** The distribution function, $f(w_\parallel|r)$, of the longitudinal component of the relative velocity measured from pairs of points in the SCDM simulation. *a)*. $f(w_\parallel|r)$ as a function of the longitudinal component of the velocity difference, $w_\parallel$. The various curves correspond to $f(w_\parallel|r)$ at separations $r$ corresponding to the clustering amplitude $\xi(r) = 50, 10, 1, 0.5,$ and $0.1$ as labeled in the figure itself. *b)*. The same curves shown in *a)* scaled by their mean $\langle w_\parallel \rangle$ and variance $\sigma$. The heavy solid curve denotes the mean of the distributions at different separations. Note the linear scale on the abscissa.

Figure 7*b* shows the $\xi(\pi)$ plots for these three models; the principal effect of changing the normalization of $\sigma$ is to put an offset into $Q(s)$, and to change the slope of $\xi(\pi)$. Figures 7*c* and *d* show $Q(s)$ and $\xi(\pi)$ as the normalization of $v_{12}(r)$ is halved and doubled. It affects the slope of $Q(s)$, but $\xi(\pi)$ is almost independent of the amplitude of $v_{12}(r)$, which means that it gives a clean measure of the velocity dispersion, as it was designed to do.

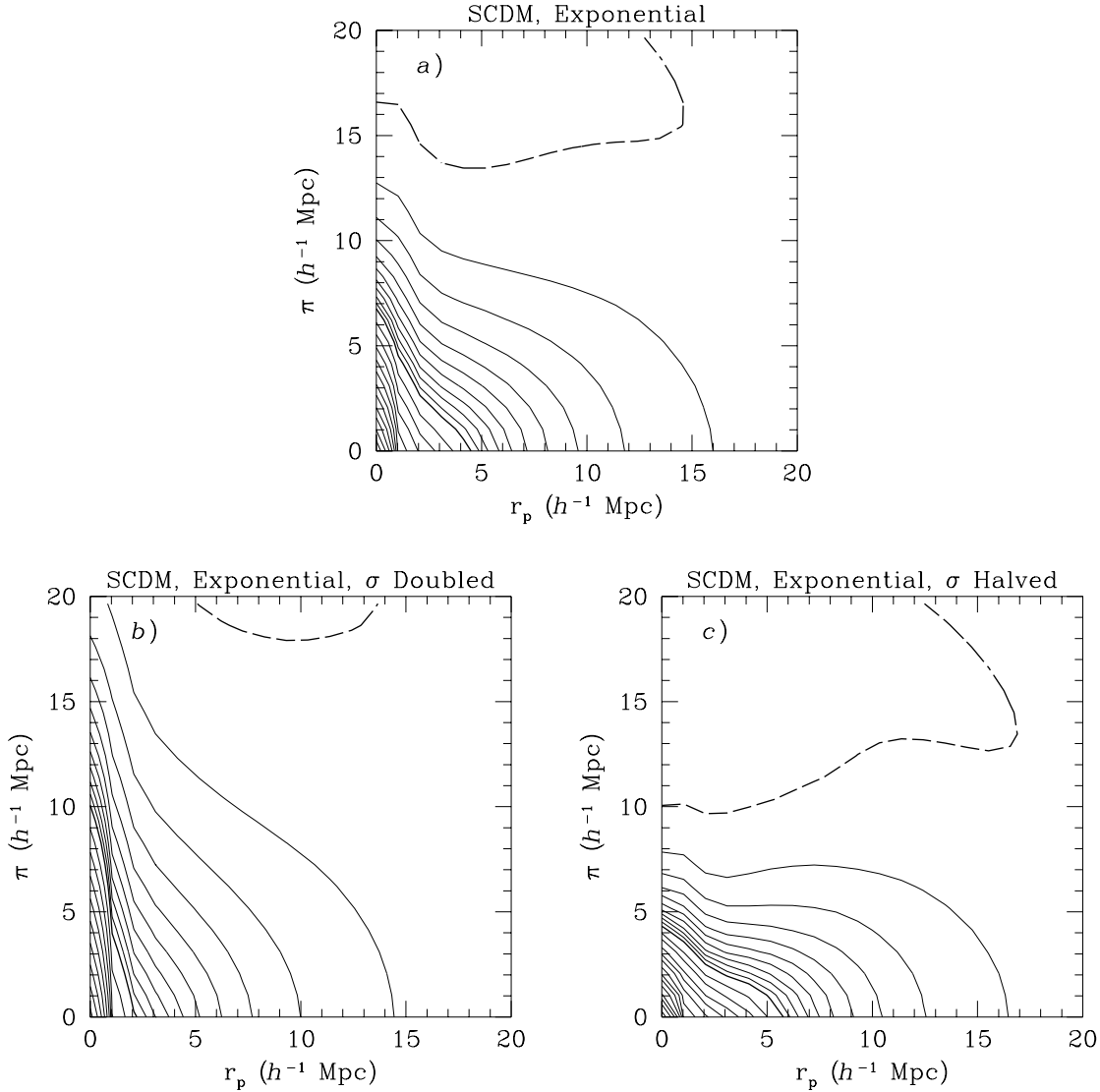

**Figure 6.** Model $\xi(r_p, \pi)$ derived using the exponential velocity distribution and the velocity moments of the unbiased SCDM simulation. Panel *a*) shows the resulting model using the velocity moments has they appear in Figure 3. In panel *b*) the amplitude of the dispersion has been doubled while in panel *c*) the amplitude has been reduced by a factor of two.

### 4.2 The Errors in $\xi(r_p, \pi)$ and Mock *IRAS* Catalogues

We now have all the ingredients necessary to make models to compare with the observed $\xi(r_p, \pi)$: the real space correlation function (Equation 4), a family of models for the velocity distribution (Equation 11), and an empirical determination from $N$-body simulations of the scaling of the first and second moments of the velocity distribution with separation (Figure 3). In order to *fit* these models to the data, however, we need to assess the statistical errors in $\xi(r_p, \pi)$. In Paper 1, we used bootstrap methods to assess the covariance matrix of the redshift space correlation function, although it became clear there that the bootstrap technique systematically overestimates the correlation function errors (c.f., Mo, Jing, & Börner 1992; Press *et al.* 1992). In addition, Figure 1 dramatically shows that the number of coherent structures included in the *IRAS* volume is small enough to leave their signatures in $\xi(r_p, \pi)$. These same structures will be present in all bootstrap resampling realizations, and thus the derived error will not reflect the uncertainty introduced by these structures. With this in mind, we take a different approach to defining the covariance matrix of $\xi(r_p, \pi)$. From the SCDM $N$-body realization, we create 100 realizations of the *IRAS* redshift survey, in the manner described in Paper 1. In brief, for each one, we choose an $N$-body point with peculiar velocity, local density, and local shear similar to those of the Local Group, and select objects around it with number density as a function of distance as given by the *IRAS* selection function. The $\xi(r_p, \pi)$ is calculated for each realization, and the covariance matrix **M** of the determination of $\xi$ at a grid of points in the $(r_p, \pi)$ plane is found from the ensemble of realizations, as described in the Appendix to Paper 1. Of course, the volume of the $N$-body simulation itself is finite, and the same structure will appear in many of the realizations. However, it will appear with a different *orientation* in each one, and thus its coupling with the redshift space distortions will differ as well. Thus we feel justified in claiming the individual realizations to be independent.

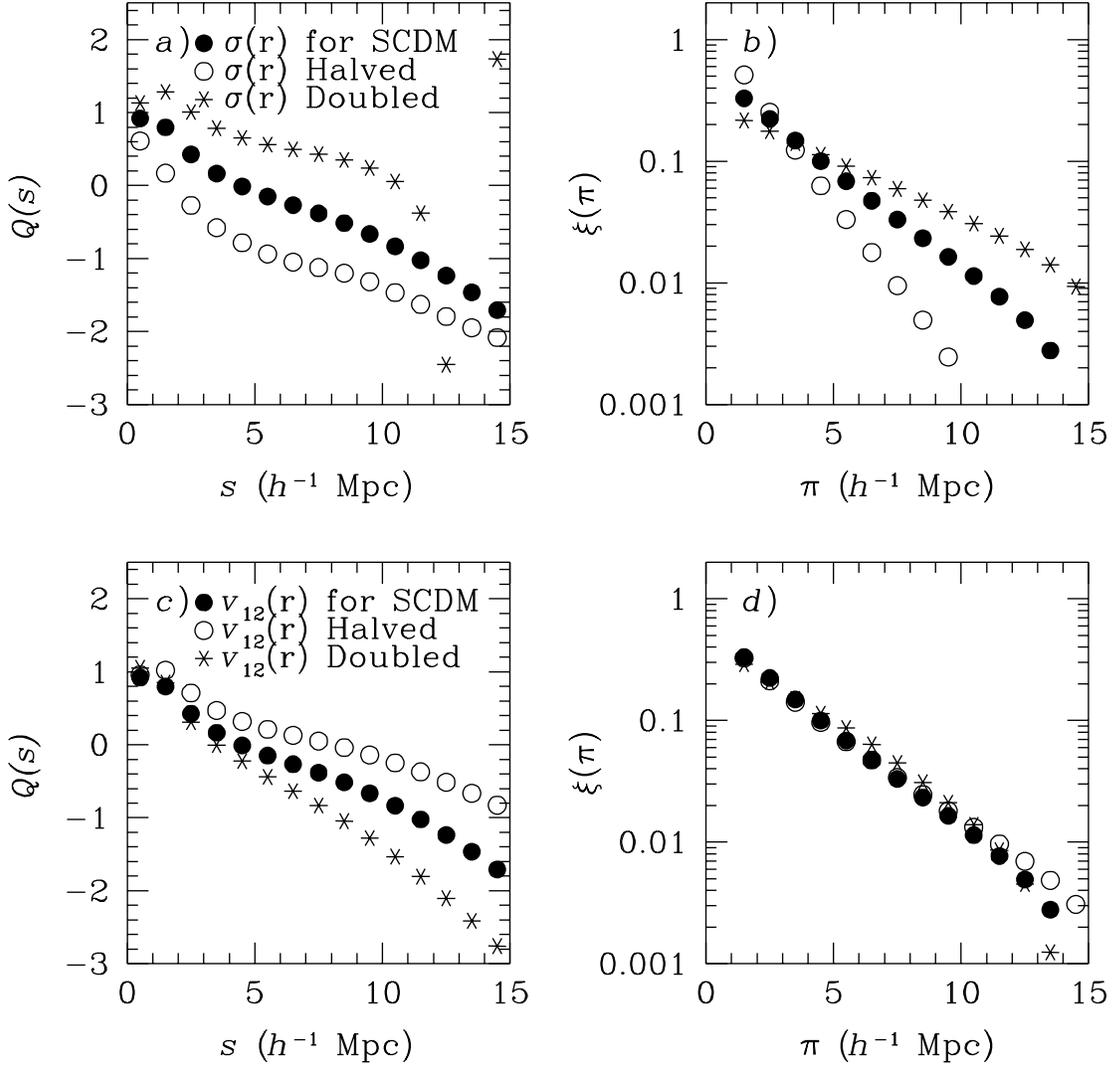

**Figure 7.** *a*). The statistic $Q(s)$ derived from the isotropic exponential model. The solid points are with the correct normalization for $\sigma(r)$, while the open circles and stars are for $\sigma(r)$ halved and doubled, respectively. These correspond to the three models of Figure 6. *b*). The statistic $\xi(\pi)$ derived from the three models in Figure 6 and panel *a*. *c*). The statistic $Q(s)$ derived from the isotropic exponential model. The solid points are with the correct normalization for $v_{12}(r)$, while the open circles and stars are for $v_{12}(r)$ halved and doubled, respectively. *d*). The statistic $\xi(\pi)$ derived from the three models in panel *c*.

Before using these realizations to determine the covariance matrix, we can compare them with models. Figure 6a above showed a model for $\xi(r_p, \pi)$, using velocity moments correctly normalized for the SCDM model. Figure 8a shows the average over the 100 realizations of $\xi(r_p, \pi)$. Figures 6a and 8a should agree if the model assumptions that went into Figure 6a were correct; indeed, Figure 8b shows that the differences between these two are substantial only in the region immediately around the origin, where the correlations are strong. However, there are two physical effects we have ignored in our model. First, the true velocity distribution function differs from the exponential assumed above, as Figure 5 shows us; there is non-negligible skewness, whose effect will be to increase the oblateness in $\xi(r_p, \pi)$ on intermediate scales ($0.5\,r_0 \lesssim r \lesssim 2\,r_0$). The approximate universal velocity distribution function given in Figure 5b can be included directly in the model. Second, the velocity dispersion tensor is anisotropic; that is, the second central moment of the longitudinal component of velocity is larger than $2^{-1/2}$ times that for the transverse component. For the $N$-body models, we can measure this anisotropy, and put it directly into the model.

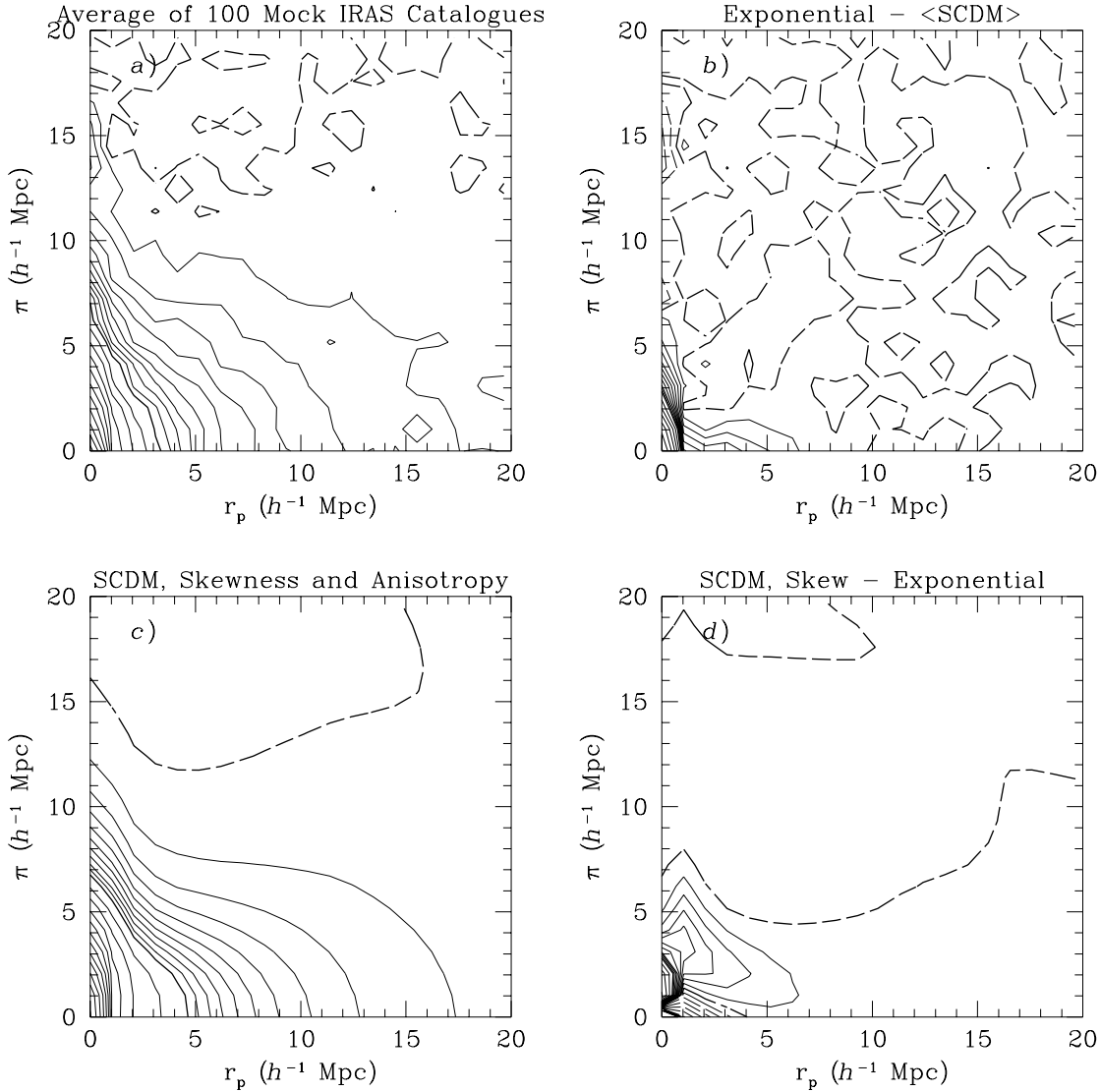

**Figure 8.** *a*). Average $\xi(r_p, \pi)$ map for the 100 mock *IRAS* SCDM catalogues. Compare with the model in Figure 6*a*. *b*). The difference between the model in 6*a* and the average of the realizations in panel *a*). All substantial differences are close to the origin. *c*). Model $\xi(r_p, \pi)$ obtained by using a spline fit to the unbiased SCDM velocity distribution function, $f(w_3|r)$ (the heavy solid curve in Figure 5*b*). In addition, the anisotropy of the velocity dispersion tensor was explicitly taken into account. Compare with Figures 6*a* and panel *a*. *d*). The difference between the anisotropic skew model in panel *b* and the isotropic exponential model in Figure 6*a*.

This tends to increase the prolateness of $\xi(r_p, \pi)$ on small and intermediate scales. Thus these two effects work in opposite senses, and when both are included in the modeling, we get the model $\xi(r_p, \pi)$ shown in Figure 8*c*. This model is quite close to the simpler model in Figure 6*a*. The difference between these two models (Figure 8*d*) is somewhat greater than that between the isotropic exponential model and the mean of the 100 mock *IRAS* catalogues; that is, the anisotropic skew model is a worse fit, even though it includes more relevant physical effects. We suspect that this is because this model overestimates the true skewness for the reasons discussed in reference to Figure 5. Although we do not show it here, adding either the refinement of skewness or anisotropy alone to the model causes a much larger change in $\xi(r_p, \pi)$ than putting both in; that is, they largely cancel one another's affects.

As we show next, the exponential model with isotropic velocity dispersion does give unbiased results for the amplitudes of the velocity moments, and thus we will use it as our primary model to compare with the data.

### 4.3 Fitting Models to the Mock Catalogues

Given the covariance matrix **M**, we define the $\chi^2$ of the difference between the data and a given model:

$$\chi^2 = \sum_{i,j} \left( \xi(r_p,\pi)_{data,i} - \xi(r_p,\pi)_{model,i} \right) M_{ij}^{-1} \left( \xi(r_p,\pi)_{data,j} - \xi(r_p,\pi)_{model,j} \right) \quad , \tag{27}$$

where the $i$ and $j$ range over the individual points in the $(r_p, \pi)$ plane at which $\xi(r_p, \pi)$ has been measured. Our first approach was to minimize the quantity $\chi^2$ with respect to the normalizations of the $\sigma(r)$ and $v_{12}(r)$ curves, for a given shape of these curves and a given form of the velocity distribution function. There are, however, several problems with this direct approach. The covariance matrix is difficult to measure accurately, even with 100 realizations, with the consequence that the estimated M can be close to singular. Furthermore, the error distribution of the 100 individual realizations shows significant skewness over all of the $(r_p, \pi)$ plane, while the $\chi^2$ statistic above is meaningful only when the error distribution is a multi-variate Gaussian. In addition, direct fits to the $\xi(r_p, \pi)$ map are necessarily very sensitive to the amplitude and slope of $\xi(r)$, and to small deviations from a power law, while we are most interested in fitting the *distortions* of the $\xi(r_p, \pi)$ contours. Presumably because of these problems, we are unable to recover the correct normalization to the velocity moments for the mock *IRAS* catalogs (in particular, our estimate of the normalization of $v_{12}$ is biased low by $\sim 20\%$).

With this in mind, we carry out our fit not to the full $\xi(r_p, \pi)$ contour map, but to the two projections we defined above in §2.2. As we described there, $\xi(\pi)$ is a statistic that is almost entirely determined by the amplitude of $\sigma(r)$, while $Q(s)$, although sensitive to $\sigma(r)$, contains most of the available information on $v_{12}(r)$ (recall Figure 7). Moreover, both statistics are normalized in such a way as to minimize their sensitivity to $\xi(r)$ itself. We find the statistics of $Q(s)$ and $\xi(\pi)$ to be well behaved over the 100 mock *IRAS* catalogues, with skewness of the error distribution consistent with zero on all scales, giving us hope that fitting the analogue of Equation 27 to the data will give unbiased results.

In fitting models to data (either the actual or mock *IRAS* catalogues), we evaluate $\xi(\pi)$ every $1\,h^{-1}$ Mpc from 0.5 to $14.5\,h^{-1}$ Mpc, and evaluate $Q(s)$ every $1\,h^{-1}$ Mpc from 1.5 to $15.5\,h^{-1}$ Mpc. The offset arises because it is not possible to measure $Q(s)$ with our binned data on smaller scale, and for convenience we want to keep the vector lengths identical. The covariance matrix of these 30 quantities is determined from the same 100 mock *IRAS* catalogues discussed above. The quantity $\chi^2$ is determined using Equation 27 above with $\xi(r_p, \pi)$ replaced with the vector containing $Q(s)$ and $\xi(\pi)$. Actually, because of the normalization applied to $\xi(\pi)$ (Equation 7), the covariance matrix is close to being singular and the calculation of the inverse of the covariance matrix in Equation 27 is unstable. We circumvent this difficulty by using using Principal Component Analysis, as described in the Appendix of Paper 1. We find linear combinations of the data points which have a diagonal covariance matrix. These linear combinations are simply the original data vector multiplied by the matrix whose columns are the eigenvectors of the original covariance matrix. Furthermore, the diagonal elements of the covariance matrix of the new data points are merely the eigenvalues of the original covariance matrix. The near singularity in the covariance matrix is manifested as one eigenvalue which is much smaller in magnitude than the others. We discard the linear combination of data points which gives rise to this eigenvalue, thereby making the fitting procedure robust. Rejection of the smallest eigenvalue reduces the number of degrees of freedom in the fit by one for a total of 27 (30 data points - 2 parameters - 1 rejected).

As a prelude to fitting models to the actual *IRAS* data, we can check the validity of our procedure using the mock *IRAS* catalogues. We calculate a grid of $\chi^2$ values as a function of the normalization of $\sigma(r)$ and $v_{12}(r)$, for each of the 100 mock *IRAS* catalogues. Here we use an exponential model for $f(w_3|r)$, which we saw above accurately reproduces the mean $\xi(r_p, \pi)$ of the mock catalogues, we use the $\xi(r)$ of the SCDM model (as measured from the $N$-body model itself), and we use the shapes of $\sigma(r)$ and $v_{12}(r)$ of SCDM (the solid curves in Figure 3 above, without correction for the virial theorem on small scales). The results are shown in the first entry of Table 1. We express the normalizations of the velocity moment curves in terms of the velocity dispersion at $1\,h^{-1}$ Mpc, $\sigma(1)$, and the streaming at $4\,h^{-1}$ Mpc, $v_{12}(4)$. For each of the 100 realizations we minimized $\chi^2$ to find the best amplitude for the velocity moments; the mean of the best fit parameters are 323.6 km s$^{-1}$ and 441.1 km s$^{-1}$, for $\sigma(1)$ and $v_{12}(4)$ respectively, within 7% of the correct answer (which we know a priori from the $N$-body model) of 349.1, 420.3 km s$^{-1}$. The *rms* scatter from realization to realization of these normalizations are 59 and 42 km s$^{-1}$, respectively, and the covariance between these quantities is small. The mean $\xi(\pi)$ and $Q(s)$ over the 100 mock catalogues are shown as the points in Figure 9, with error bars given by the standard deviation over these catalogues. The solid curve is the best fit isotropic exponential model. The other curves will be discussed below.

Figure 10 shows the cumulative distribution of the derived values of $\chi^2$ for the hundred mock *IRAS* catalogues, for the best fit for each model (solid histogram, mean value 32.3) and when $\sigma(r)$ and $v_{12}(r)$ are held at their correct values (dotted histogram, mean value 34.7). If our model is a correct description of the data, the solid histogram should agree with the $\chi^2$ distribution for 27 degrees of freedom, which is shown as the light smooth curve. They are clearly in poor agreement; the true distribution is better fit (although not perfectly!) for 32.3 degrees of freedom (heavy smooth curve), implying that our model is not a perfect match to the data. We have repeated this experiment limiting the fits to the range $1-10\,h^{-1}$ Mpc in $Q(s)$ and $\xi(\pi)$, as the noise gets very large beyond this range; the excess in the mean $\chi^2$ over the expected value remains, although the derived $v_{12}(4)$ and $\sigma(1)$ are in agreement with those listed in Table 1.

Thus our model is not a perfect descriptor of the $N$-body data; our average $\chi^2$ is larger than the number of degrees of freedom by five, meaning that there are physical effects which we have not taken into account. Nevertheless, we have demonstrated that minimizing $\chi^2$ gives unbiased results for the $\sigma(r)$ and $v_{12}(r)$ normalization. This is reassuring, because we know the model for the velocity distribution function is not correct in detail, since we have ignored its skewness and the anisotropy of the velocity dispersion tensor. However, we saw above that the effects of these two refinements on $\xi(r_p, \pi)$ approximately cancel each other, and ignoring them does not introduce any strong bias in the derived parameters. In fact, our effort modeling them greatly worsens the fit. As mentioned above, this is probably due to our overestimation of the true skewness; we have not attempted to refine the these anisotropic models, and we drop them from further attention. Because of the relatively poor signal to noise ratio of an individual realization, the simple isotropic exponential model for the velocity distribution is more than adequate, and will serve as our preferred model for the available *IRAS* data. As much larger redshift surveys become available, it may become desirable to refine our model by including additional effects within the velocity distribution function.

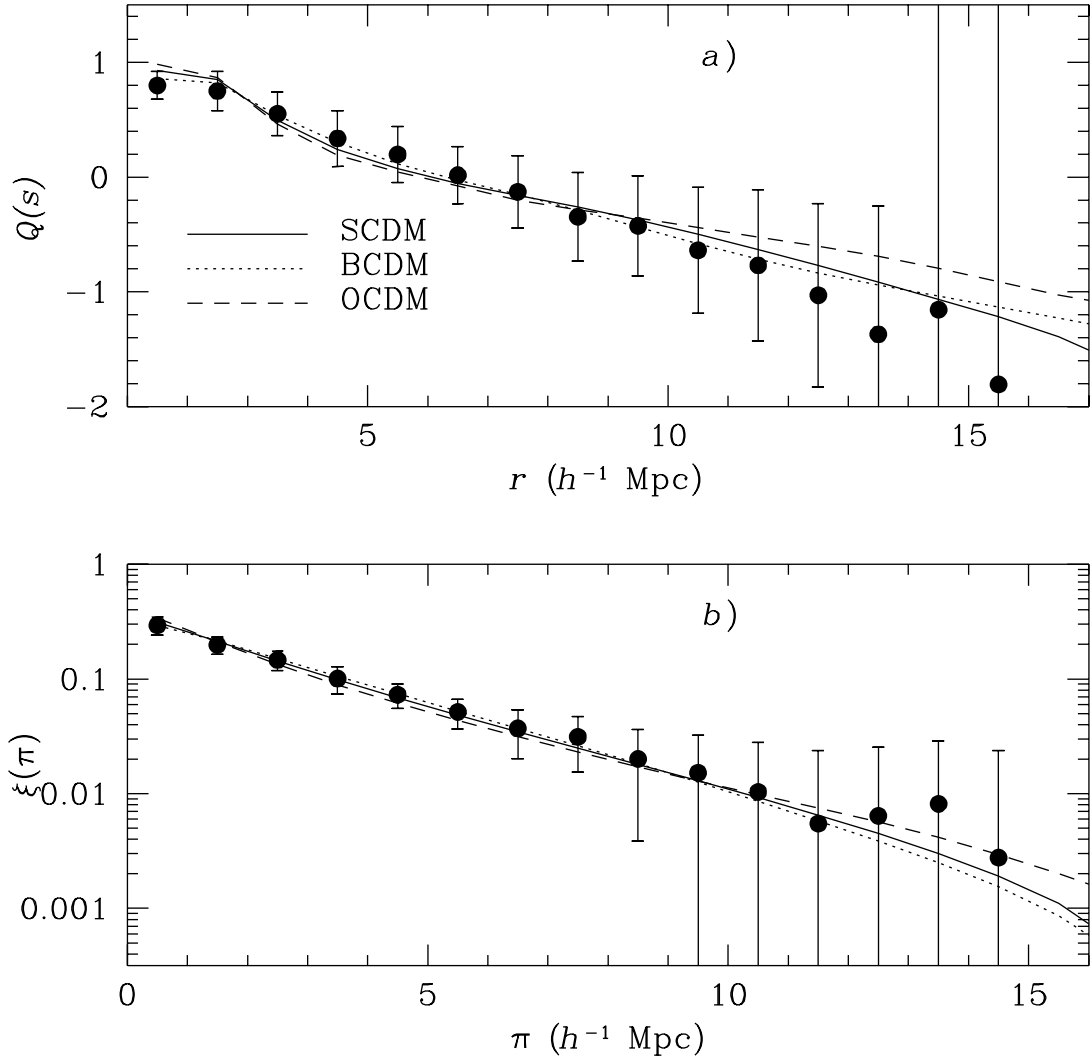

**Figure 9.** *a*). The solid points plot $Q(s)$ for the average of the 100 mock catalogues, while the error bars show the dispersion around this mean. The curves show the mean of the best fit models to each of these catalogues as given in Table 1, for various forms of the velocity distribution functions. *b*). The same as in panel *a*, now for the $\xi(\pi)$ statistic.

Thus far, we have shown that when we use the correct functional form of $\sigma(r)$ and $v_{12}(r)$ for the simulation, we can recover their correct normalizations. Of course we do not know the functional form of the velocity moments for the *IRAS* data. Therefore our ability to model real data rests on the insensitivity of our model $\xi(r_p, \pi)$ to the exact functional form of $\sigma(r)$ and $v_{12}(r)$. We saw in Figure 3 that the scale dependence of the velocity moments for different cosmological models was similar although not identical. As a test we have tried to recover the normalization of the velocity moments of the SCDM mock *IRAS* catalogs with various other choices for the model of the velocity moments and distribution functions. Table 1 also shows results of fitting the 100 mock catalogues to both a Gaussian ($\nu = 2$ in Equation 12) and hybrid ($\nu = 1.5$) velocity distribution function and to the velocity moments from the open (OCDM) and biased (BCDM) simulations. Also shown in Table 1 is the results for the SCDM model with both the skewness of the velocity distribution ($f(w_3|r)$ given by the spline fit in Figure 5) and the anisotropy of the velocity dispersion taken into account; we refer to this model as ACDM.

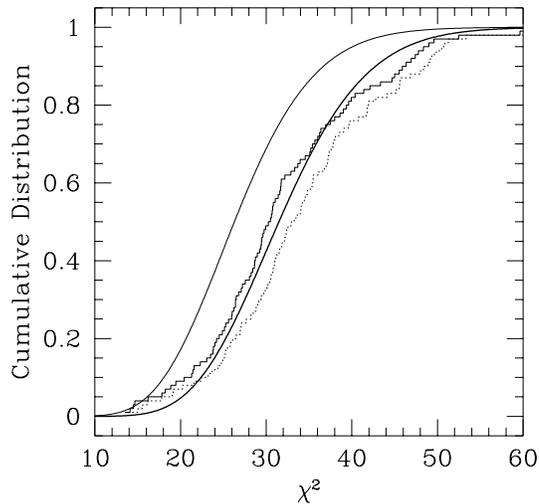

**Figure 10.** Cumulative distribution of the best $\chi^2$ (solid histogram) and the $\chi^2$ with normalization fixed at the correct value (dotted histogram) from fits (using an exponential velocity distribution function and the SCDM velocity moments) to $Q(s)$ and $\xi(\pi)$ of the 100 mock *IRAS* catalogs. The light and heavy smooth curves are the expected $\chi^2$ distributions with 27 and 32.3 degrees of freedom, respectively.

The mean $\chi^2$ values show that the exponential model is greatly favored over the other models for the velocity distribution function. This is because $\xi(\pi)$ gives us such a clear signature of what the correct functional form of $f(w_3)$ must be. Indeed, the exponential model is preferred over the spline fit of Figure 5, presumably for the reasons discussed in reference to that figure. The mean of the derived velocity dispersion at $1\,h^{-1}\,\mathrm{Mpc}$ is quite insensitive to the model used, giving results close to the right answer. However, the derived streaming at $4\,h^{-1}\,\mathrm{Mpc}$ does depend sensitively on both the shape of the velocity distribution function, and more ominously, on the form of the velocity moments used. Figure 9 shows that although the normalizations of the velocity distribution functions are quite different in the different cases, the model fits to $Q(s)$ and $\xi(\pi)$ are almost indistinguishable from one another, and are in good agreement with the data. In addition, when we fit models to the *IRAS* data, we will apply the modifications to the velocity moments discussed in § 4.1 above, bringing them into closer agreement with each other and thus reducing the discrepancy in normalization between them. Finally, the linear theory expression for $v_{12}$ of course only holds on large scales; note that the normalization found for the BCDM and SCDM models (the two with the best $\chi^2$) agree much better at $10\,h^{-1}\,\mathrm{Mpc}$ than at $4\,h^{-1}\,\mathrm{Mpc}$. Figure 4 shows that with the correct normalization, $v_{12}$ does indeed yield a correct estimate of $\beta$, giving us confidence that we will derive an unbiased estimate of $\beta$ from the *IRAS* data themselves.

## 5  MODELING THE *IRAS* $\xi(r_p, \pi)$

Having developed statistical procedures that give meaningful answers when tested on $N$-body models, we are ready to apply them to the *IRAS* data themselves. We have fit series of models to the whole sky *IRAS* data, the *IRAS* data with clusters boosted and to the Northern and Southern subsamples, with results listed in Table 2. For each fit, the 1-$\sigma$ errors on the two normalizations are given, as calculated from the projection of the $\Delta\chi^2 = 1$ contour. In addition, we give the value of $v_{12}(r)$ at $r = 10\,h^{-1}\,\mathrm{Mpc}$, a scale that is fully in the linear regime, as Figure 4 shows. Finally, we list $\chi^2_{min}$, the value of $\chi^2$ for the best fit model. The fits to the North and South subsamples used covariance matrices calculated from 100 mock *IRAS* catalogues covering half the sky.

Figure 11 shows the $\xi(\pi)$ and $Q(s)$ curves for the full-sky sample, together with the expected 1-$\sigma$ error bar per point from the mock *IRAS* catalogues. The three almost indistinguishable model curves shown are from the best-fit normalization of the SCDM, BCDM, and OCDM models with an exponential velocity distribution function. The models agree with one another better than was found for the fits to the $N$-body models, because the current fits use velocity moment shapes that have been modified to be in close agreement with one another (compare the left and right panels of Figure 3). Indeed, Table 2 shows nearly identical results for the normalizations derived from the various models.

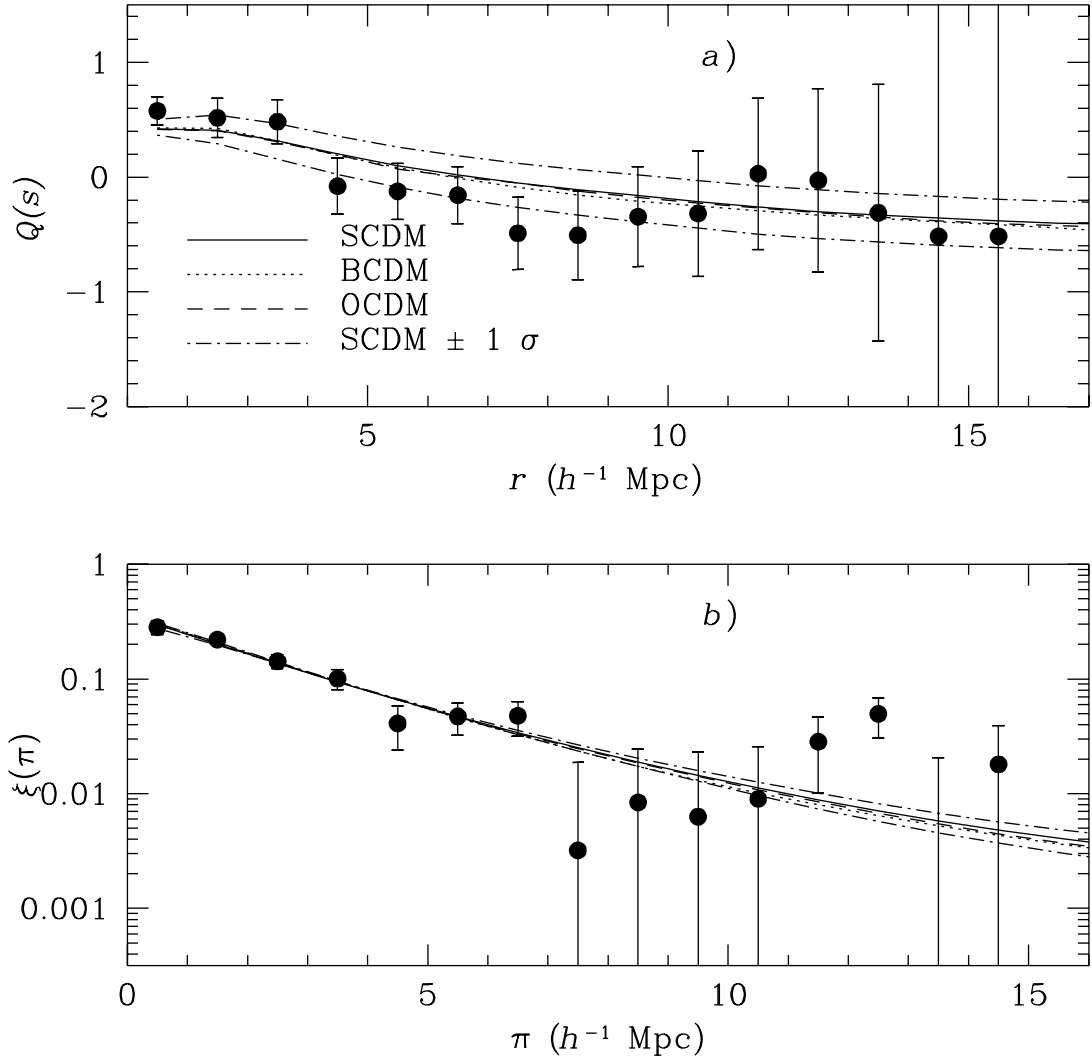

**Figure 11.** *a*). Plot of $Q(s)$ for the *IRAS* whole sky sample. The error bars are the standard deviation over the 100 SCDM mock *IRAS* catalogues, and are thus the same as are shown in Figure 9. The curves are the best fit normalizations given in Table 2 for various models. In addition, we show the model curves when the normalization of $v_{12}$ is moved up and down by its one-$\sigma$ errors (in the SCDM exponential fit). *b*). Same as in panel *a*, for the $\xi(\pi)$ statistic.

The two flanking dot-dashed curves in Figure 11 give the results of the SCDM model when the best-fit value of $v_{12}(4)$ is raised and lowered by its 1-$\sigma$ errors. The best-fit model cannot go through both the first set and second set of three points in the $Q(s)$ plot, and strikes a compromise between the two. When $v_{12}(4)$ is raised and lowered by its errors, one obtains curves that go through one or the other of these sets.

Table 2 shows that the minimum values of $\chi^2$ for our fits agree with each other quite well. This means that we cannot distinguish between the different models for the velocity moments, at least with the modifications made to them to bring them into closer agreement with the expected forms for the *IRAS* galaxies. However, the minimum values of $\chi^2$ are somewhat lower than we found with the mock *IRAS* catalogs, even when we used exactly the right form for the correlation function and velocity moments in the models. Indeed, Figure 10 shows that only 13% of the mock *IRAS* catalogues give a $\chi^2 \leq 22.0$. The SCDM model has a correlation function which is substantially steeper than that of *IRAS* galaxies, which means that the covariance matrix may have overestimated the errors of the observed correlations on small scales. Although this is unlikely to strongly bias the best-fit models (it is hard to see how the model fits in Figure 11 could be improved) it will bias the derived value of $\chi^2$ down, causing us to overestimate the errors on derived parameters slightly.

An example of the constraint in parameter space $(\sigma(1), v_{12}(4))$ from the models is given in Figure 12, for fits to the isotropic exponential model using the SCDM velocity moments. The innermost contour describes the $1-\sigma$ error limits for each parameter taken separately (giving rise to the errors listed in Table 2). The second contour encloses 68% of the likelihood, while the third contour describes the $2-\sigma$ error limits for each parameter. There is strong covariance between the parameters, although $\sigma(1)$ is better constrained than is $v_{12}(4)$.

Figure 1 shows that giving extra weight to clusters in the *IRAS* sample has only modest effects on $\xi(r_p, \pi)$. This is borne out by the model fits; the boosted sample shows slightly higher velocity dispersion than the unboosted case. This is reassuring; because the velocity dispersion is a pair-weighted statistic, we worried that our results would be quite sensitive to the way in which clusters are treated. It is possible, of course, that the galaxies in the very core of rich clusters, where the velocity dispersion is highest, are missing altogether from the *IRAS* sample, in which case no amount of cluster boosting would bring them back. However, Strauss *et al.* (1992a) found that the Finger of God of the Virgo cluster in *IRAS* and optically-selected galaxies had roughly the same extent in the redshift direction, arguing against this possibility.

The velocity dispersion found in the South subsample is in good agreement with that for the full sky, while that for the North is somewhat larger, as we saw already from the $\xi(r_p, \pi)$ maps themselves in Figure 1. However, in both subsamples, the estimate of $v_{12}(4)$ is much higher than that in the full sky, even though the $Q(s)$ statistics are in good agreement between the different subsamples (Figure 2); rather than choosing the compromise solution of the full sky that goes between the first six points of Figure 11a, the best-fit solution for the North and South subsamples goes through the second set of three points (like the lower dot-dashed curve of Figure 11a). However, the errors in $v_{12}(4)$ (at least in the North) are enormous, and include the value found for the full sky. From Figure 2 it is clear that there is appreciable noise in the estimates of $Q(s)$ and $\xi(\pi)$ even in the full sample; cutting the sample in half only amplifies the noise. Moreover, the elements of the covariance matrix for the half-sky mock catalogs are considerably larger simply because of counting statistics. These two effects combine to make the $\chi^2$ contours for the velocity moments in the Northern and Southern subsamples very broad, and the derived normalizations of $v_{12}(r)$ and $Q(s)$ subject to large fluctuations.

We conclude that the *IRAS* data are best fit by

$$\sigma(1) = 317^{+40}_{-49} \text{ km s}^{-1},$$
$$v_{12}(4) = 167^{+99}_{-73} \text{ km s}^{-1}, \quad (28)$$
$$v_{12}(10) = 109^{+64}_{-47} \text{ km s}^{-1}.$$

The velocity dispersion estimate is consistent with the value derived for the CfA1 survey, $340 \pm 40$ km s$^{-1}$ (Davis & Peebles 1983), again implying that the undercounting of cluster cores by *IRAS* galaxies is not a serious issue. The streaming result has not previously been reported.

The SCDM model corresponding to the parameters in Equation 28 is shown in Figure 13b. This should be compared with the *IRAS* $\xi(r_p, \pi)$ from the full sample in Figure 13a (copied from Figure 1a). The difference of the two is shown as Figure 13c. The model is not perfect, particularly on small scales, but does provide a reasonable fit to the distortions on scales $\gtrsim 3\,h^{-1}$ Mpc.

We have assumed the virialized form for $\sigma(r)$ on small scales for all analyses of the *IRAS* data. It is sobering to see the effect of dropping this assumption. We took two extreme models; in the first, we used the velocity moments of the SCDM model in which no smoothing on small scales had been done; so that $\sigma(r)$ *rises* on small scales. In this case, the oblateness of the $\xi(r_p, \pi)$ contours at large $r$ is largely driven by velocity dispersion rather than infall (compare Figure 7), and the data are consistent with no infall at all (i.e., $v_{12}(4) = 0$)! Alternatively, we use the SCDM model for $v_{12}(r)$, but for $\sigma(r)$ we use a form that grows more slowly than the virialized form:

$$\sigma(r) = \begin{cases} \sigma(1) r^{1/2} & \text{if } r \leq 10\,h^{-1}\text{Mpc}; \\ \sigma(1)(10)^{1/2} & \text{if } r > 10\,h^{-1}\text{Mpc}. \end{cases} \quad (29)$$

In this case, the best fit values of $v_{12}(4)$ is now 417km s$^{-1}$, more than twice the previous value. In both of these models, the best value of $\sigma(1)$ was essentially identical with that found above; moreover, all models gave comparable best $\chi^2$ values. We have no a priori reason to believe that $\sigma(r)$ differs substantially from a virialized form on small scales (indeed, Davis & Peebles (1983) present evidence that $\sigma(r)$ is a slowly rising function of $r$ in small scales, close to the virialized form), but if it does, our estimate of $v_{12}(4)$ could be in error by a large factor. Our estimate of $\sigma(1)$ is quite robust to such uncertainties.

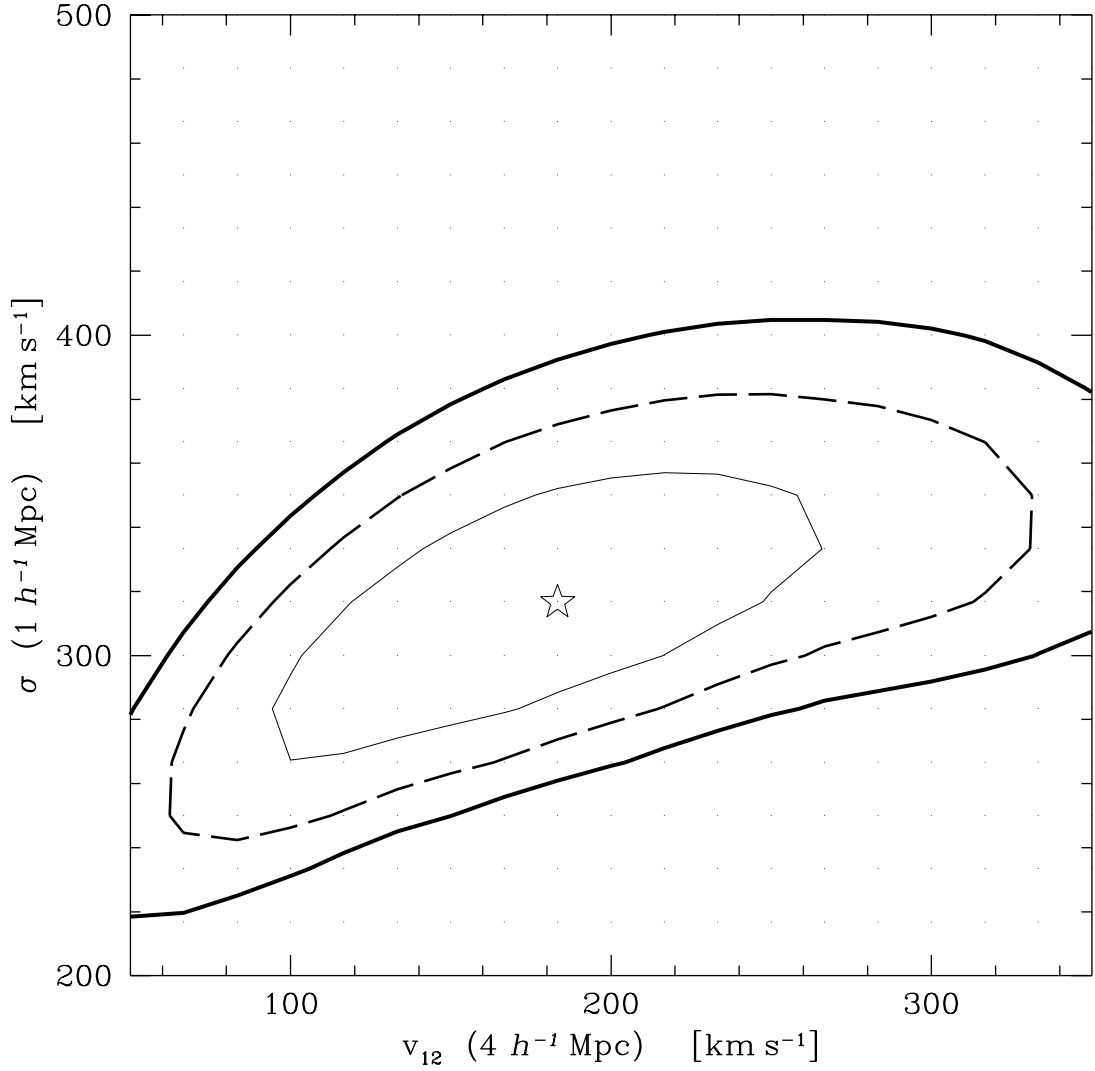

**Figure 12.** Contours at $\chi^2$ larger than the minimum value (marked with a star) by 1.0, 2.3, and 4.0. These correspond to the $1-\sigma$ confidence level for the parameters taken separately, the 68% joint confidence interval, and the $2-\sigma$ confidence level for the parameters taken separately. These are based on the isotropic exponential fit to the whole sky *IRAS* data, using the SCDM model curves. The dots indicate the grid of values at which models were calculated.

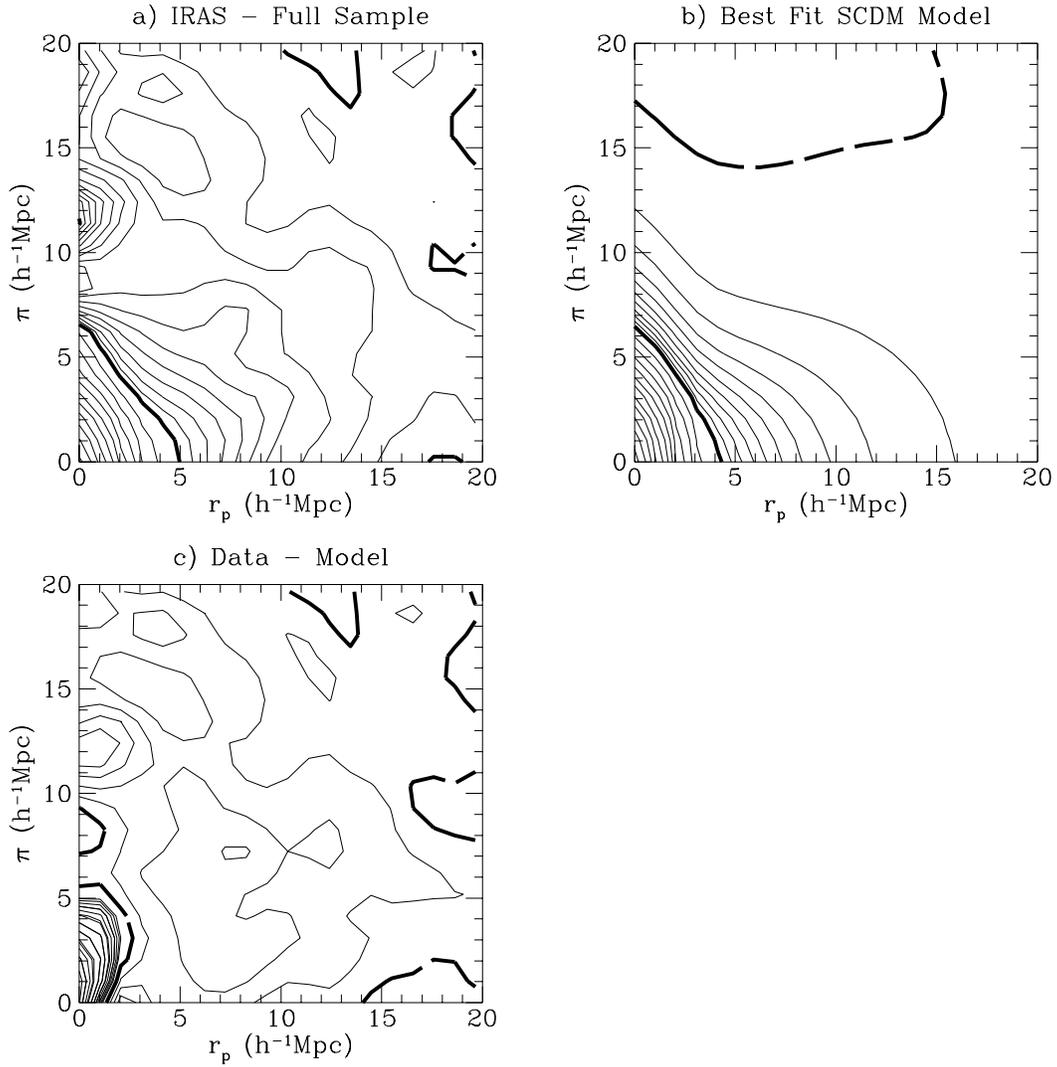

**Figure 13.** *a*). The *IRAS* 1.2 Jy $\xi(r_p, \pi)$ for the full sample as in panel *a* of Figure 1. *b*). The best fit SCDM model corresponding to the parameters in Equation 28 with an isotropic exponential velocity distribution function. *c*.) The difference between the *IRAS* and SCDM model. Contour levels are the same as in Figure 1. For clarity, the maps have been twice smoothed by a 1-2-1 boxcar in each direction.

## 6 DISCUSSION

In § 3 we described how measurement of the redshift distortions of the $\xi(r_p, \pi)$ maps provides two independent estimators of the cosmic density parameter $\Omega$. Both of these estimates are sensitive to bias relating the galaxy and mass distributions, but their dependency is different in the two cases.

In order to apply the Cosmic Virial Theorem, we need a value for $Q$ which expresses the ratio between the second and third moments of the *IRAS* galaxy density field. Bouchet *et al.* (1993), using a counts in cells analysis, derived a value for the "normalized" skewness, $S_3 = 1.5 \pm 0.5$, which is related to $Q$ by $Q = \frac{1}{3} S_3 J_3 / J_2^2$ where $J_2$ and $J_3$ are dimensionless integrals defined by Peebles (1980). For the *IRAS* value of $\gamma = 1.66$ this yields $Q = 0.48 \pm 0.16$. This is roughly half the value obtained for optically selected galaxy samples (e.g., Groth & Peebles 1977), and is a reflection of the relative undercounting of cluster cores in *IRAS* selected galaxy samples (Strauss *et al.* 1992a). Indeed, Bouchet *et al.* found that the derived value of $S_3$ (and hence $Q$) is extremely sensitive to the way in which clusters are treated in the analysis, and that giving them extra weight increased $Q$ by a factor of two.

The $J(\gamma)$ integral which appears in Equation 24 can be evaluated numerically for the *IRAS* value of $\gamma = 1.66$ to yield 4.28. On small scales, $\Sigma^2 \gg v_{12}^2$, thus the latter term can be dropped from Equation 24. Because we have used the Cosmic Virial Theorem to force the shape of $\sigma(r)$ at small distances, our estimate for $\Omega/b$ is independent of $r$ from 1 to 3.76 $h^{-1}$ Mpc. Thus we find $\Omega/b = 0.38$. What is the error in this quantity? Statistical errors arise in the determination of $\sigma$, $S_3$, $\gamma$, and $r_o$, which lead to an estimated $1 - \sigma$ error in $\Omega/b$ of $\pm 0.24$. Large as this statistical error is, it is overshadowed by systematic effects. The derived value of $Q$ from the *IRAS* sample is exquisitely sensitive to the treatment of clusters; boosting them causes $Q$ to double (Bouchet *et al.* 1993). Deviations of the two and three point correlation functions from their assumed power law forms will lead to further systematic error.[‡] The Cosmic Virial Theorem is derived assuming statistical equilibrium of close pairs, and requires taking the continuum limit (Peebles 1980), neither of which can be justified observationally. Finally, as discussed earlier, it is only with perfect linear bias that the factor '$b$' within the Cosmic Virial Theorem holds, although linear bias becomes an oxymoron on the highly non-linear scales on which the Cosmic Virial Theorem is valid (cf., Szalay 1988). We attempted to apply the Cosmic Virial Theorem to the $N$-body models themselves with little success; the correlation functions differed too strongly from power laws to get meaningful results. Thus we conclude that we are unable to get a meaningful estimate of cosmological parameters from the Cosmic Virial Theorem, and in fact do not believe that the assumptions which go into it (perfect power law correlation functions, perfect linear biasing, isotropy of the velocity dispersion, the continuum limit, and statistical equilibrium of clustering) have been properly validated for any extant data sample.

Although we cannot use our measurement of $\sigma(1)$ to obtain a useful estimate of $\Omega$, its value does put strong constraints on cosmological models. Standard CDM seems to require strong biasing to be consistent with the small value observed (Davis *et al.* 1985; Gelb & Bertschinger 1993), although it has been argued that if galaxies have systematically smaller velocities than do dark matter particles, the model can be saved (Couchman & Carlberg 1992). Recently, models with a mixture of hot and cold dark matter have become popular; these hybrid models better match the large scale constraints and at the same time have somewhat lower amplitude velocity fields on smaller scales, but it is not yet clear if their pair dispersion is an adequate match to the observations (Klypin *et al.* 1993; Davis, Summers, & Schlegel 1992).

A more reliable estimate of cosmological parameters uses our measurement of $v_{12}(r)$ on large scales (Equation 21). Figure 4 above showed that given the correct amplitude for $v_{12}(r)$, this method indeed yields the correct value of $\beta$. Figure 14 shows the derived value of $\beta$ as a function of scale from the normalization of $v_{12}$ and our various models for the run of $v_{12}$ with distance. Because we have scaled $v_{12}(r)$ to agree with the linear theory prediction given the *IRAS* correlation function, the curves must asymptote to a constant value on large scales. However, the fact that the curves using different models largely agree is significant, and gives us confidence in the reality of the result. The curves give $\beta = 0.45^{+0.27}_{-0.18}$ $(1 - \sigma)$; the statistical error on this is dominated by that of $v_{12}(4)$. We do not include the possible systematic error due to uncertainty in the shape of $\sigma(r)$ on small scales (cf., the discussion at the end of the previous section).

This measurement of $\beta$ is on a length scale of 5-10 $h^{-1}$ Mpc, and is consistent either with the *IRAS* galaxies having a bias $b = 2.2^{+1.5}_{-0.8}$ in an Einstein-de Sitter Universe, or with the *IRAS* galaxies being an unbiased tracer in an $\Omega = 0.26^{+0.32}_{-0.15}$ universe. From Table 2, we note that the Southern and Northern hemispheres alone yield estimates of $v_{12}(4)$ considerably higher than found for the full sky analysis. These estimates are consistent with $\beta = 1$, but with considerably larger error than for the full sky sample. Although all subsamples are consistent with each other within $1.5\sigma$, they span an enormous range of derived $\beta$, indicating that the present redshift catalogs are too small for a definitive application of this statistic.

Note that the allowed range of $\beta$ lies between the typical values found on cluster and smaller scales from virial arguments, $\Omega/b \sim 0.1$ (or $\beta \sim 0.25$ if $b = 1$), (Faber & Gallagher 1979; White 1990 and references therein) and the best fit value of $\beta = 1.28 \pm 0.3$ $(1\sigma)$ given by the *POTENT-IRAS* comparison on a scale of 15-40 $h^{-1}$ Mpc (Dekel *et al.* 1993; see also Kaiser *et al.* 1991; Strauss *et al.* 1992c; Fisher, Scharf, & Lahav 1993). If all these observations are correct, then we apparently live in a high density universe in which the bias in the galaxy distribution decreases strongly with scale. Scale dependent bias operating on scales of less than 1 $h^{-1}$ Mpc, where baryonic cooling and dynamical friction are effective, would not be too surprising (White & Rees 1978; Cen & Ostriker 1992; Couchman and Carlberg 1992). However, extending this scale dependence to larger scales is difficult to understand. Only more precise estimates of $v_{12}(r)$, derived from much larger redshift surveys, can improve our error bars substantially.

---

[‡] The Cosmic Virial Theorem can be generalized for any form of the two and three-point correlation functions, but this would require knowledge not only of $S_3$, but the full form of $\zeta(\mathbf{x}, \mathbf{r}, \mathbf{x} - \mathbf{r})$, which has not been calculated for *IRAS* galaxies.

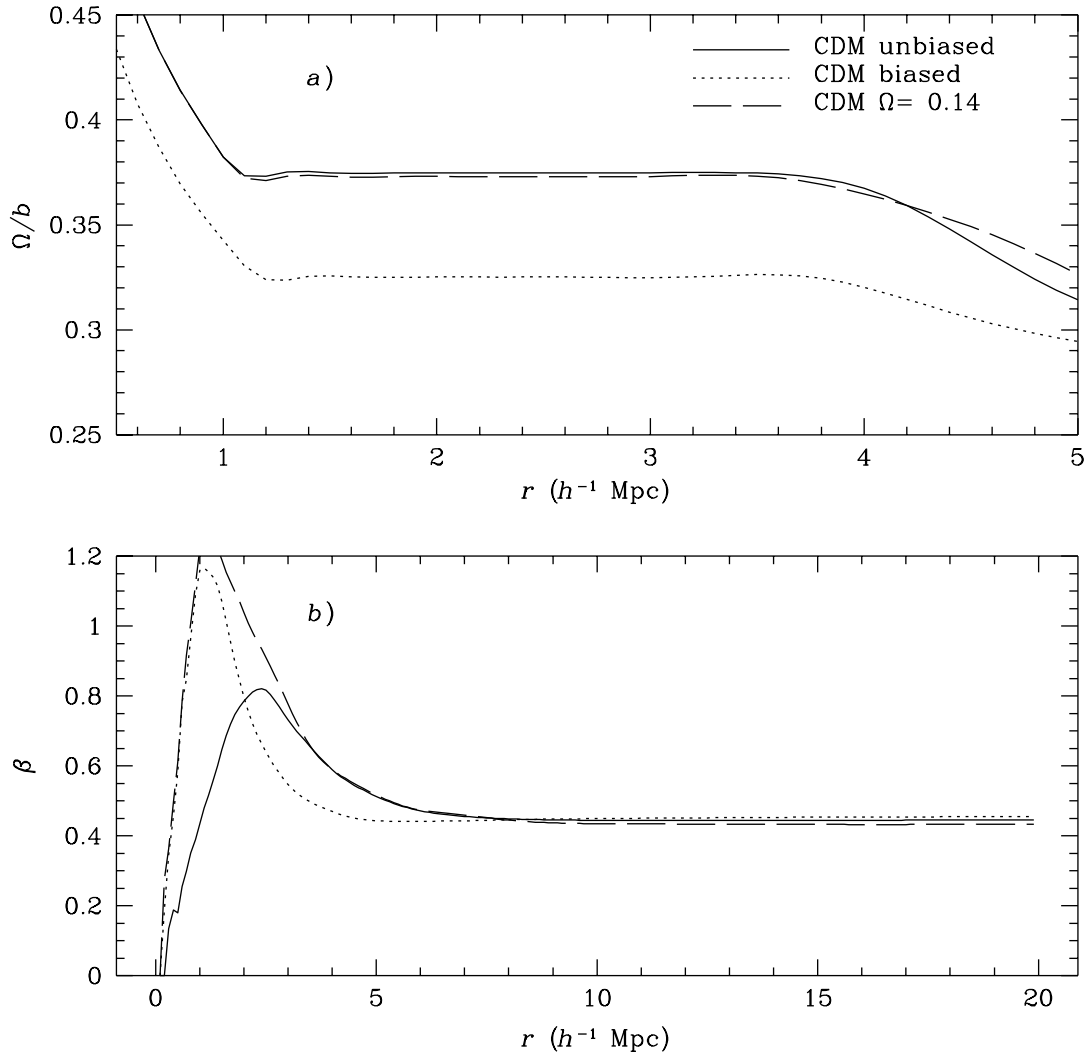

**Figure 14.** Constraints on $\Omega$ and $b$ from the derived velocity dispersion and large scale streaming. The results for different models for the velocity moments are shown with different line types. a). Estimates of $\Omega/b$ from the Cosmic Virial Theorem and the derived normalization to $\sigma(r)$ for the *IRAS* full-sky data. Because we use the Cosmic Virial Theorem to modify the $N$-body $\sigma(r)$ on small scales, the results are independent of scale from 1 to $3.76\,h^{-1}$ Mpc. As discussed in the text, systematic errors make this determination worthless. b). Estimates of $\beta$ from Equation 21 and the derived normalization to $v_{12}(r)$ for the *IRAS* full-sky data. Because we use linear theory to modify the $N$-body $v_{12}(r)$ on large scales, the results are independent of scale above $\sim 7\,h^{-1}$ Mpc.

As discussed above, our estimate of $v_{12}(r)$ will have large systematic error if our assumed form of $\sigma(r)$ is grossly in error. The velocity dispersion differs greatly from its linear theory limit on all scales that we probe, thus assuming linear theory alone will cause *systematic* errors in any derivation of $v_{12}$. This interdependency of the infall streaming and the *nonlinear* pair dispersion is inevitable, and in our assessment effectively invalidates Hamilton's (1993a) analysis which is based solely on the effects of linear theory. Hamilton extends his analysis to extremely large scales, but the expected infall streaming is very small on these scales, and as Figure 6 demonstrates, on all scales for which one can hope to measure the distortions in $\xi(r_p, \pi)$, the role of the non-linear component of $\sigma(r)$ cannot be neglected.

In order to reconcile an apparent large value of $v_{12}(4)$ with a low density universe, it would be necessary to argue that the true $\sigma(r)$ is a strongly declining function of scale (much more so than shown in the OCDM simulation), and we have erred in modeling it as very shallow. It is difficult to see how this could be, particularly since the low density models of structure formation eschew bias in the galaxy distribution, and the measured $\sigma(r)$ in this model must exist at roughly the inferred value merely to satisfy the Cosmic Energy Equation and balance the potential energy stored in the galaxy fluctuations.

Conversely, to accommodate a high density unbiased universe with our measured streaming, one must argue that $\sigma(r)$ grows more steeply with scale than has been modeled in our analysis. The strong covariance between $\sigma(r)$ and $v_{12}(r)$ might then permit a larger estimate for the inferred streaming, which we presently estimate to be more than a factor of two smaller than expected in an $\Omega = 1$, unbiased universe (i.e. $\approx 2\sigma$ from $\beta = 1$). Our virialization procedure for $\sigma(r)$ on small scales insures $\sigma(4)/\sigma(1) = 4^{1-\gamma/2} = 1.27$. Given the strong constraint on $\sigma(1)$, a larger asymptotic value of $\sigma(r > r_0)$ would imply either a steeper gradient in $\sigma(r)$, or a larger scale for the cutoff, or both. It seems dubious that sufficient power in the matter distribution on the scale of 5-10 $h^{-1}$ Mpc exists to generate this excess dispersion. Power on still larger scale (such as to provide for large amplitude flows on scales greater than 3000 km s$^{-1}$) is ineffective at generating relative dispersion between pairs of galaxies on the scales under consideration here.

## 7 SUMMARY

The main conclusions of this paper can be summarized as follows:

- Redshift distortions on small and large scales are observed unambiguously in the 1.2 Jy *IRAS* galaxy redshift survey. These provide a statistical measure of peculiar velocities, obtained purely from redshifts, with no independent measurements of distance.
- The redshift correlation function $\xi(r_p, \pi)$ can be modeled adequately on scales $\lesssim 15$ $h^{-1}$ Mpc by a convolution of the real space correlations with the relative velocity distribution function. We can infer information about the first two moments of this velocity distribution. The velocity distribution function itself can be taken to be an isotropic exponential; skewness and anisotropy of the distribution function create distortions of opposite sign and roughly cancel one another. Much larger redshift surveys, such as the Sloan Digital Sky Survey (Gunn & Knapp 1993) will have the signal-to-noise ratio to allow us to look for these more subtle physical effects. This survey will cover a much greater volume than that of the *IRAS* survey, greatly decreasing the effect of individual structures in $\xi(r_p, \pi)$, and thus allowing a much cleaner measure of the redshift distortions. Modeling such higher quality data will require $N$-body simulations which reproduce the correlation function of the survey more closely than the simulations used in this paper, eliminating the need to rescale the functional forms of the velocity moments, and reducing the possibility of systematic errors in the covariance matrix.
- The mean relative streaming of *IRAS* galaxies on a scale of $4\,h^{-1}$ Mpc is $v_{12}(4) = 167^{+99}_{-73}$ km s$^{-1}$ while the pairwise velocity dispersion at $1\,h^{-1}$ Mpc is $\sigma(1) = 317^{+40}_{-49}$ km s$^{-1}$. This streaming flow acts to cancel the Hubble flow; our measurements imply approximately half of the Hubble expansion of pairs is canceled on the scale of $r_0 \approx 4\,h^{-1}$ Mpc. The mean streaming measurement has not previously been reported, while the pairwise velocity dispersion is consistent with previous estimates derived from optically selected redshift catalogues. The results are robust to double-counting cluster cores, but the amplitude of the mean streaming has high covariance with the derived pair dispersion. Moreover, it is quite sensitive to the model assumed for the dispersion on small scales.
- The Cosmic Virial Theorem applied to the derived *IRAS* velocity dispersion leads to an estimate of $\Omega/b \sim 0.38 \pm 0.24$. The relatively large uncertainty in this estimate reflects the statistical uncertainties of a nearly divergent integral over the three point correlation function. However, systematic errors are expected to be even more severe, and largely invalidate the Cosmic Virial Theorem as an estimator of cosmological parameters.
- Linear theory and the derived estimate of the streaming give an estimate of $\beta = \Omega^{0.6}/b = 0.45^{+0.27}_{-0.18}$ (statistical errors). The derived amplitude of the streaming is sensitive to the form assumed for the velocity dispersion on small scales. Substantial improvement of the error bars on this estimate will require much larger redshift surveys.
- Our inferred $\beta$ is applicable to an intermediate scale, $\sim 10 - 15\,h^{-1}$ Mpc, and we have obtained results intermediate between the lower values determined from estimates of cluster masses $\Omega/b = 0.1$, $r < 1\,h^{-1}$ Mpc), and those obtained on much larger scale ($\beta = 1.28$, $15 < r < 40\,h^{-1}$ Mpc). If all these results are correct, they argue for strong scale dependence of galaxy biasing.

**Acknowledgments** We thank Simon White, Shaun Cole, Adi Nusser, Will Sutherland, David Weinberg, and Ben Moore for useful discussions and Guinevere Kauffmann for providing the velocity moments of the OCDM model. KBF acknowledges a Berkeley Department of Education Fellowship and a SERC postdoctoral fellowship. MAS is supported at the IAS under NSF grant # PHY-9245317, and grants from the W.M. Keck Foundation and the Ambrose Monell Foundation. The work of MD was partly supported by NSF grant AST-9221540 and NASA grant NAG-51360.

**TABLE 1**

**Fits of 100 Mock *IRAS* SCDM Catalogues to Models**

| Moments | $f(w_3)$ | $\langle v_{12}(4)\rangle$ | Error | $\langle v_{12}(10)\rangle$ | Error | $\langle \sigma(1)\rangle$ | Error | $\langle \chi^2_{min}\rangle$ | $\langle \chi^2_{true}\rangle$ |
|---|---|---|---|---|---|---|---|---|---|
| SCDM | Exponential | 323.6 | 59.4 | 175.6 | 32.2 | 441.1 | 42.4 | 32.3 | 34.7 |
| SCDM | Gaussian | 407 | 78.5 | 220.9 | 42.6 | 339.3 | 22.2 | 44.4 | 57.7 |
| SCDM | $\nu = 1.5$ | 354.6 | 66.6 | 192.5 | 36.1 | 360.1 | 28.6 | 37.6 | 43.4 |
| BCDM | Exponential | 244.0 | 55.2 | 166.5 | 37.7 | 457.4 | 50.0 | 29.4 | 37.3 |
| OCDM | Exponential | 415.0 | 71.6 | 282.8 | 48.8 | 407.9 | 40.8 | 38.6 | 41.3 |
| ACDM | Exponential | 239.3 | 57.5 | 129.9 | 31.2 | 441.5 | 44.1 | 33.3 | 38.9 |
| ACDM | Spline | 248.3 | 102.9 | 134.7 | 55.8 | 393.9 | 55.5 | 56.7 | 60.5 |

Means are over the 100 SCDM Mock *IRAS* catalogues. $v_{12}(4)$, $v_{12}(10)$, $\sigma(1)$, and their errors (i.e., the standard deviations over the realizations) are listed in km s$^{-1}$. The two $\chi^2$ columns are means of the best solution, and means of the solution with the correct values of $v_{12}(4)$ and $\sigma(1)$ imposed. The correct values of $v_{12}(4)$, $v_{12}(10)$, and $\sigma(1)$ are 349.1, 189.5, and 420.3 km s$^{-1}$, respectively.

**TABLE 2**

$Q(s)$, $\xi(\pi)$ **Fits of** *IRAS* **Data to Models**

| Moments | $f(w_3)$ | $v_{12}(4)$ | $v_{12}(10)$ | $\sigma(1)$ | $\chi^2_{min}$ |
|---|---|---|---|---|---|
| | | Fits to Full Sky | | | |
| SCDM | Exponential | $167^{+99}_{-73}$ | $109^{+64}_{-47}$ | $317^{+40}_{-49}$ | 22.0 |
| BCDM | Exponential | $133^{+60}_{-48}$ | $111^{+50}_{-40}$ | $300^{+40}_{-33}$ | 21.3 |
| OCDM | Exponential | $167^{+81}_{-61}$ | $107^{+52}_{-39}$ | $317^{+35}_{-36}$ | 22.2 |
| | | Fits to Full Sky, Clusters Boosted | | | |
| SCDM | Exponential | $167^{+74}_{-72}$ | $109^{+48}_{-47}$ | $333^{+44}_{-46}$ | 15.8 |
| BCDM | Exponential | $133^{+42}_{-59}$ | $111^{+35}_{-49}$ | $317^{+45}_{-38}$ | 15.3 |
| OCDM | Exponential | $150^{+58}_{-61}$ | $96^{+37}_{-39}$ | $333^{+40}_{-40}$ | 16.0 |
| | | Fits to Northern Sky | | | |
| SCDM | Exponential | $333^{+56}_{-183}$ | $218^{+37}_{-120}$ | $367^{+54}_{-47}$ | 21.3 |
| BCDM | Exponential | $217^{+42}_{-75}$ | $180^{+35}_{-62}$ | $350^{+47}_{-51}$ | 20.5 |
| OCDM | Exponential | $250^{+48}_{-42}$ | $160^{+31}_{-27}$ | $333^{+59}_{-37}$ | 20.1 |
| | | Fits to Southern Sky | | | |
| SCDM | Exponential | $317^{+72}_{-85}$ | $207^{+47}_{-55}$ | $317^{+50}_{-36}$ | 13.1 |
| BCDM | Exponential | $217^{+52}_{-40}$ | $180^{+43}_{-33}$ | $300^{+38}_{-46}$ | 12.5 |
| OCDM | Exponential | $250^{+43}_{-42}$ | $160^{+27}_{-27}$ | $300^{+36}_{-45}$ | 12.3 |

All errors given are $1-\sigma$.